\documentclass[prb,twocolumn,aps,groupedaddress,showpacs,amssymb,prb,floatfix,preprintnumbers]{revtex4}
\usepackage{graphicx,color}
\usepackage{subfigure}
\usepackage{verbatim}
\definecolor{drkgr}{rgb}{0.05,0.5,0.2}

\begin{document}
\thispagestyle{myheadings}
\title{The magnetic phase diagram of the frustrated spin chain compound linarite, PbCuSO$_4$(OH)$_2$, as seen by neutron diffraction and $^1$H-NMR}

\author{L. Heinze$^1$, G. Bastien$^2$, B. Ryll$^{3}$, J.-U. Hoffmann$^3$, M. Reehuis$^3$, B. Ouladdiaf$^4$, F. Bert$^5$, E. Kermarrec$^5$, P. Mendels$^5$, S. Nishimoto$^{2,6}$, S.-L. Drechsler$^2$, U. K. R{\"o}{\ss}ler$^2$, H. Rosner$^7$, B. B\"uchner$^{2,8}$, A. J. Studer$^9$, K. C. Rule$^9$, S. S\"{u}llow$^1$, and A. U. B. Wolter$^2$}
\affiliation{$^1$Institut f\"{u}r Physik der Kondensierten Materie, TU Braunschweig, D-38106 Braunschweig, Germany\\
$^2$Leibniz-Institut f\"{u}r Festk\"{o}rper- und Werkstoffforschung IFW Dresden, D-01171 Dresden, Germany\\
$^3$Helmholtz-Zentrum Berlin f\"{u}r Materialien und Energie, D-14109 Berlin, Germany \\
$^4$Institute Laue-Langevin, F-38042 Grenoble Cedex, France \\
$^5$Laboratoire de Physique des Solides, CNRS, Univ. Paris-Sud, Universit\'e Paris-Saclay, F-91405 Orsay Cedex, France\\
$^6$Institut f\"ur Theoretische Physik, Technische Universit\"at Dresden, D-01068 Dresden, Germany \\
$^7$Max-Planck-Institut f\"ur Chemische Physik fester Stoffe, D-01068 Dresden, Germany \\
$^8$Institut f\"ur Festk\"orper- und Materialphysik, Technische Universit\"at Dresden, D-01062 Dresden, Germany \\
$^9$Australian Centre for Neutron Scattering, ANSTO, Kirrawee DC, New South Wales 2234, Australia}
\date{\today}

\begin{abstract}
We report on a detailed neutron diffraction and $^1$H-NMR study on the frustrated spin-1/2 chain material linarite, PbCuSO$_4$(OH)$_2$, where competing ferromagnetic nearest neighbor and antiferromagnetic next-nearest neighbor interactions lead to frustration. From the magnetic Bragg peak intensity studied down to 60\,mK, the magnetic moment per Cu atom is obtained within the whole magnetic phase diagram for $H \parallel b$ axis. Further, we establish the detailed configurations of the shift of the SDW propagation vector in phase V with field and temperature. Finally, combining our neutron diffraction results with those from a low-temperature/high-field NMR study we find an even more complex phase diagram close to the quasi-saturation field suggesting that bound two-magnon excitations are the lowest energy excitations close to and in the quasi-saturation regime. Qualitatively and semi-quantitatively, we relate such behavior to $XYZ$ exchange anisotropy and contributions from the Dzyaloshinsky-Moriya interaction to affect the magnetic properties of linarite.
\end{abstract}
 
 \maketitle

\section{Introduction}

In quantum spin systems the interplay of low dimensional magnetic exchange paths, quantum fluctuations and magnetic frustration often leads to unconventional and exotic magnetic properties which have attracted attention in recent years \cite{balents2010,starykh2015,norman2016,coldea2002,azurite2008,willenberg2012,depenbrock2012,buettgen2014,weickert2016,du2016,balents2016}. As a result of this interplay, in real materials conventional long range magnetic order can be suppressed down to very low temperatures, possibly leading to novel magnetic states such as spin liquids \cite{balents2010, norman2016,depenbrock2012,isono2016}. As well, they may display a variety of exotic in-field behavior such as unusual spin density wave (SDW) or spin nematic phases \cite{willenberg2012,buettgen2014,weickert2016,balents2016,vekua2007,sudan2009,heidrich-meissner2010,zhitomirskyLiCuVO4,nawa2013LiCuVO4,starykh2014,nawa2014NaCu,onishi2015,pregelj2015,smerald2016,grafe2017}. 

In particular, regarding the latter issue, one essential model, that has been studied intensively, is the one-dimensional spin-1/2 chain, where frustration occurs due to competing nearest neighbor and next-nearest neighbor magnetic exchange. Correspondingly, this model is described by the Hamiltonian
\begin{equation}
\label{Model1}
\mathcal{H} = J_1 \sum_i \mathbf{S}_i\,\mathbf{S}_{i+1} + J_2 \sum_i \mathbf{S}_i\,\mathbf{S}_{i+2} - h \sum_i S_i^z,
\end{equation}
with $\mathbf{S}_i$ being the spin-1/2 operator at chain site $i$. The parameters $J_1$ and $J_2$ correspond to the nearest neighbor (NN) and next-nearest neighbor (NNN) interaction between spins $\mathbf{S}_i$, $\mathbf{S}_{i+1}$ and $\mathbf{S}_{i+2}$, respectively. For this model, complex phase diagrams in applied magnetic fields have been predicted, including the above mentioned spin density wave and spin-multipolar phases. The appearance of these states sensitively depends on the frustration ratio $\alpha = J_2 / J_1$ \cite{balents2016,vekua2007,sudan2009,zhitomirskyLiCuVO4,starykh2014,onishi2015,smerald2016}. Experimental studies on materials such as LiCuVO$_4$, LiCu$_2$O$_2$, PbCuSO$_4$(OH)$_2$, LiCuSbO$_4$, $\beta$-TeVO$_4$ or NaCuMoO$_4$(OH) \cite{willenberg2012,buettgen2014,weickert2016,nawa2013LiCuVO4,nawa2014NaCu,pregelj2015,hikihara2008,sato2009,sato2013,dutton2012,saul2014,nawa2015} have verified some of these predictions, while neither a full experimental characterization of the various in-field phases nor understanding of the observed phenomena has been achieved.

Conceptually, the materials considered as model compounds for the frustrated $J_1$-$J_2$ chain belong to the class of edge-sharing Cu$^{2+}$ or V$^{4+}$ (both carrying spin $S = 1/2$) chain systems. Here, because of a nearest-neighbor magnetic exchange $J_1$ between spins on a chain provided by an oxygen bond with a bond angle close to 90$^{\circ}$ and a next-nearest neighbor exchange $J_2$ via a stretched O--O bond, $J_1$ may be ferro- or antiferromagnetic, while $J_2$ is antiferromagnetic, leading to frustration between $J_1$ and $J_2$. 

For most of the materials considered as realizations of the frustrated $J_1$-$J_2$ chain, experimental studies are hampered because (large) single crystals are often lacking, crystallographic disorder affects the magnetic properties, high fields are needed to reach saturation~\cite{Schumann2018}, or a combination of all these factors. In this respect, the natural mineral linarite, PbCuSO$_4$(OH)$_2$, which has been modeled as a frustrated isotropic spin-1/2 chain system \cite{effenberger,kamieniarz2002,schofield2009,wolter2012,willenberg2012,schaepers2013,willenberg2013,schapersthesis,Schaepers2014,willenberg2016,povarov2016,rule2017,Mack2017,Cemal2018,Feng2018}, appears to be the most accessible for experiments. 

Linarite crystallizes in a monoclinic structure (space group: $P2_1/m$)~\cite{effenberger,schofield2009,schaepers2013} of buckled, edge-sharing Cu(OH)$_2$ units aligned along the crystallographic $b$ axis. Modeling with linear spin wave theory indicates that the exchange parameters most likely are $J_1 = -114$\,K and $J_2 = 37$\,K corresponding well to values observed from bulk property measurements~\cite{rule2017}. A consistent model also requires an interchain interaction $J_{\rm interchain} = 4$\,K, {\it i.e.}, $\sim$\,5\% of $J_1$. Its quasi-saturation field~\cite{remsat} of the order of 10\,T allows for experiments to be performed within a very wide range of the magnetic phase diagram, which has been shown to consist of at least five regions/phases I--V for $H \parallel b$ axis (see Fig.~\ref{fig:PhaseDiagram})~\cite{willenberg2016,povarov2016,Feng2018}. Recently, phase V has been proposed to represent a longitudinal spin-density wave phase with the SDW propagation vector $\vec{q}$ shifting with the magnetic field~\cite{willenberg2016}. In addition, a phase separation between this SDW phase and a non-dipolar ordered phase was suggested in high magnetic fields~\cite{willenberg2016}. So far, the SDW phase has not been fully characterized and an alternative proposal has been put forward explaining phase V as a so-called fan state~\cite{Cemal2018}. In addition, evidence has been given that, for crystal directions away from the $b$ axis, phase V is also existent~\cite{willenberg2013,Feng2018} while the relationship of phase V to the other thermodynamic phases is not fully understood at present. 

Here, we present a combined detailed neutron diffraction and NMR study on linarite. Compared to our previous studies~\cite{willenberg2012,schaepers2013,willenberg2016}, with our present experiment we cover the magnetic phase diagram $H \parallel b$ axis up to magnetization saturation and down to lowest temperatures. This way, we substantially expand our characterization of phase V, and in particular establish the temperature and field dependence of the SDW propagation vector $\vec{q}$ beyond the limits of our previous study~\cite{willenberg2016} and far beyond the experimental characterization realized for any other $J_1$-$J_2$ chain material~\cite{mourigal2012}. From our measurements the mapping of the magnetic moment per Cu atom was derived within the whole magnetic phase diagram, revealing the nature of the transitions from phases I, III, IV and V into each other. The quasi-saturation field was precisely identified from NMR measurements performed at 160\,mK. By combining the results from neutron diffraction, the NMR spectra and NMR relaxation rate measurements a highly unusual behavior was observed at low temperatures close to the quasi-saturation field, which might be related to the realization of a spin nematic state.

\section{Experimental details}
\subsection{Neutron diffraction}

Neutron diffraction measurements were carried out at the single-crystal diffractometer D10 of the Institute Laue-Langevin in Grenoble, France, at the flat-cone diffractometer E2 of the Helmholtz-Zentrum Berlin f\"{u}r Materialien und Energie, Germany, and by using the high-intensity diffractometer WOMBAT at ANSTO, Australia. At D10, the measurements were carried out on a 26\,mg single-crystalline mineral sample (origin: Blue Bell Mine, San Bernadino, USA) which had already been used for previous neutron diffraction~\cite{willenberg2012,willenberg2016} and NMR studies~\cite{wolter2012}. The linarite crystal has a mosaic spread of a few degrees leading to broadened reflections in the neutron diffraction experiment. 

For the D10 experiment a $^3$He/$^4$He dilution cryostat was used in order to carry out measurements in the temperature range between 50\,mK and 1.5\,K (PG monochromator, neutron wavelength $\lambda = 2.36$\,\AA). The sample was aligned in an external magnetic field such that the field pointed along the crystallographic $b$ axis, with a maximum field of 10\,T. In order to perform a mapping of the neutron scattering intensity of the magnetic Bragg peaks appearing within the different magnetically ordered phases, scans in reciprocal space were carried out along (0 $k$ 1/2), ($h$ 0 1/2) and ($h$ 0.186 1/2). This way, all phases reported to be magnetically ordered were covered (see Fig.\,\ref{fig:PhaseDiagram} for the magnetic phase diagram of linarite for $H \parallel b$ axis). For the measurements, $\mu_0 H$-$T$ mesh scans were carried out. The magnetic field was set and, first, scans of (0 $k$ 1/2) with varying $k$ were performed for different temperatures in the high field region of the magnetic phase diagram, {\it i.e.}, mainly the SDW phase V and part of phase IV. The temperature was swept at constant magnetic field, starting at the lowest temperature point studied for the respective field. Next, scans in the intermediate field region (3 to 7.5\,T) were carried out for wide ranges of phase IV as well as parts of phase V. In phase IV, the commensurate magnetic Bragg peak at (0 0 1/2) was scanned along ($h$ 0 1/2) by varying $h$. In phase V, again $k$ scans were performed. Finally, phase I was investigated by $h$ scans along ($h$ 0.186 1/2).  

The linarite sample from the D10 experiment was studied in the same geometry at E2, here using a $^3$He cryostat for the intermediate temperature range (0.4 -- 2.2\,K) of the magnetic phase diagram of linarite for $H \parallel b$ axis. Further, an additional neutron diffraction experiment focused on the high temperature region of the magnetic phase diagram (temperatures $> 1.5$\,K) was carried out at E2, this time using a $^4$He cryostat in combination with a specialized sample stick with a temperature stability of 0.5\,mK. At E2, for both experiments, the maximum magnetic field was 6.5\,T aligned along the $b$ axis. Again, thermal neutrons with a neutron wavelength of $\lambda = 2.38$\,\AA ~were used.

WOMBAT was set up with the graphite monochromator for a wavelength of $\lambda = 4.22\,$\AA. The single-crystalline 27\,mg mineral sample (origin: Grand Reef Mine, Graham County, USA) was mounted in the $^3$He/$^4$He dilution stick within the 12\,T magnet. The measurements were carried out in applied fields up to 10\,T. 

At E2, the magnetic phases were studied by performing $\omega$-scans of the magnetic Bragg peaks (0 0 1/2) in phase IV and (0 $k_y$ 1/2) in phase V. For the determination of the $k_y$ value, where (0 $\pm k_y$ 1/2) is the incommensurability vector of the SDW, also the (0 0.186 1/2) peak in phase I was scanned, serving as reference point for the $k_y$ determination. Similarily, the WOMBAT measurements were conducted on a sample aligned within the ($h$ 0 $l$) scattering plane such that the incommensurate part of the wave vector (0 $k_y$ 1/2) could be observed as out of plane scattering. Independent field and temperature scans were performed. The large position sensitive detector bank of WOMBAT allowed us to observe the change in the incommensurate wave vector with applied field and temperature in phase V. Further, the hysteretic behavior of the magnetic Bragg peaks in region II below $\sim 500$\,mK in fields between $\sim 2 - 3.5$\,T were studied by field scans.

\subsection{Nuclear magnetic resonance (NMR)}

A 12.4\,mg single crystal from the Grand Reef Mine, Graham County, Arizona, USA, was used for the $^1$H-NMR experiment. The sample quality was checked by magnetization measurements and found to be in full agreement with the samples used in the neutron study. The $^1$H-NMR experiment was performed in a dilution refrigerator with temperatures from 60\,mK up to 1.1\,K in magnetic fields up to 11\,T. The NMR spectra were collected by field scans at constant frequency $\nu_0$ using a $\pi/2$-$\tau$-$\pi$ Hahn spin-echo pulse sequence. The dilution refrigerator and the sample holder contain hydrogen in amorphous or molecular form, which results in a sharp peak at zero NMR shift in the NMR spectrum, but which does not interfere with the intrinsic signal from the sample due to the NMR shift of linarite at high magnetic fields. The $1/T_1$ relaxation rate measurements were performed using the saturation recovery method.

\section{The present level of modeling and understanding of linarite}

Linarite PbCuSO$_4$(OH)$_2$ has been known as a natural mineral for more than two hundred years \cite{sowerby1809,brooken1822,glocker1839,shannon1919,chukhrov1939}, and for more than 120 years its more than 40 localities around the world have been described by mineralogists. The physical and chemical period of the linarite research history started about 70 years ago when its unit cell and the low-symmetry lattice structure became partially known step by step beginning from the 1950's \cite{berry1951}. Presently, to improve the refinement of experimental elastic and inelastic neutron scattering data larger single crystals as compared to available natural ones are highly desirable. This issue has been addressed recently by growing the first synthetic crystalline samples~\cite{rule2018}.

After the discovery of CuGeO$_3$ as a frustrated spin-Peierls system it became clear that all edge-sharing chain cuprates exhibit a non-negligible frustrating antiferromagnetic (AFM) NNN intrachain coupling $J_2$ irrespective of the sign of $J_1$, and so does linarite. Fitting susceptibility data within an isotropic $J_1$-$J_2$ model Kamenienarz {\it et al.}~\cite{kamieniarz2002} arrived at a ferromagnetic, but small $J_1$ value of $-30$\,K and a frustration ratio $\alpha = 0.5$. After the discovery of magnetic ordering below about 2.8\,K and especially the elucidation of a very complex magnetic phase diagram~\cite{willenberg2012,willenberg2016} it became clear that the knowledge of relatively weak interchain interactions and anisotropic exchange couplings is very important to understand quantitatively its details. 

After the prediction of multipolar phases at high magnetic fields slightly below saturation in frustrated $J_1$-$J_2$ chains~\cite{sudan2009,heidrich-meissner2010} the search for exotic field induced novel quantum states became one of the central issues in frustrated quantum magnetism. Thereby the important role of interchain exchange and spin anisotropy in real compounds may play a decisive role for the very existence of such states. The magnetic phase diagram of linarite displays marked anisotropies~\cite{schaepers2013,Schaepers2014,willenberg2016,povarov2016,Mack2017,Cemal2018} and requires to account for these small anisotropic interactions. The monoclinic symmetry of linarite allows for anisotropic exchange, for the bilinear intra- and interchain couplings, but also antisymmetric staggered Dzyaloshinsky-Moriya interactions (DMI) do exist, {\it e.g.} for the intrachain NN- and diagonal interchain exchange, which modify the basic spin model of Eq.~(\ref{Model1}).

At present, there is no consensus about such refined anisotropic spin models. Different simplified models have been adopted, which are difficult to compare. In Ref.~\cite{Cemal2018}, a very strong exchange anisotropy only for the NN-bonds along the chain has been hypothesized to account for the anisotropy of the spin-system in linarite alone. This proposal, however, requires specific cancellations of the DMIs which are allowed by the low symmetry and usually are expected to contribute the most important anisotropic spin-lattice coupling effects in a spin system of Cu$^{2+}$ ions. A rigorous analysis of a realistic $XYZ$ spin-model including vital DM couplings of linarite has not been performed so far due to its mathematical difficulty. 

A first step in the study of anisotropic exchange has been undertaken recently~\cite{willenberg2016} with the help of DMRG. Here, we extend this type of analysis by describing low temperature magnetization data for a model with anisotropic terms (see Fig.~\ref{fig_DMRG} and discussion in section IV.c.). High-field aspects including the role of a staggered DM interaction will be discussed in Ref.~\cite{Schumann2018}. Finally, the knowledge of these couplings including also weak interchain interactions is necessary for a theoretical analysis of the rich phase diagram especially at high magnetic fields close to the quasi-saturation, where the role of novel dipolar~\cite{Cemal2018,smerald2016} and multipolar states is currently under debate. In this context, the elucidation of the numerous very subtle couplings represents the task at hand for a deeper understanding of the properties of linarite. The experimental data given here define the framework of possible sophisticated scenarios.

\section{Results}

\subsection{Neutron diffraction}

\begin{figure}[t!]
\begin{center}
\includegraphics[width=0.95\columnwidth]{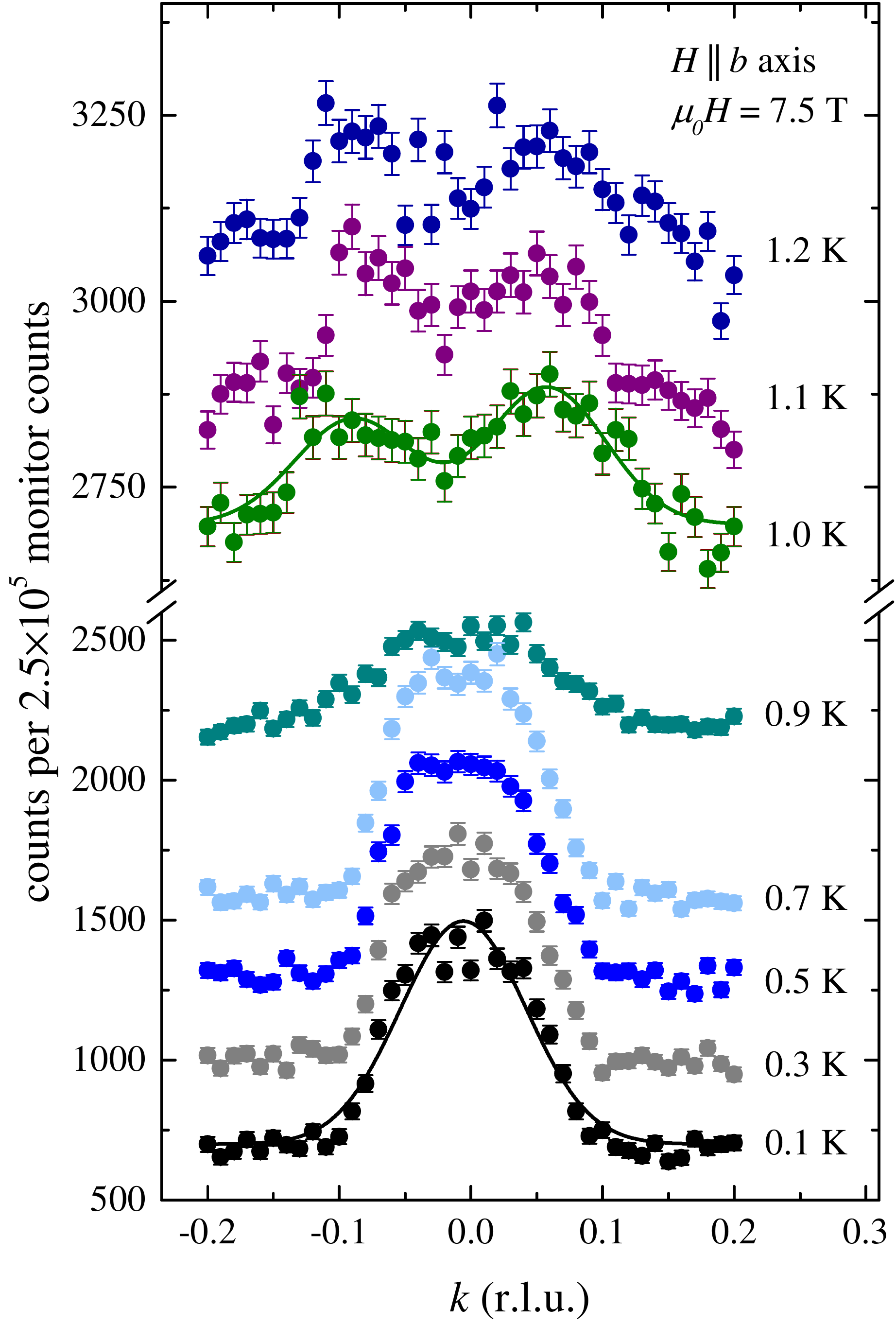}
\end{center}
\caption[1]{Scans in reciprocal space along (0 $k$ 1/2) with varying $k$ recorded for several temperatures in a field of 7.5\,T at the instrument D10, ILL. The phase boundary IV--V can be identified by the change in the appearance of the magnetic Bragg peaks. The peaks are broadened due to a slight mosaic spread within the linarite crystal. The data for $T > 0.1$\,K are shifted along the vertical axis for clarity. As example, for the scans at 0.1 and 1.0\,K a Gaussian fit curve is added to the data.} 
\label{fig:Overview_kscans_D10_7p5T}
\end{figure}

\begin{figure*}[t!]
\begin{center}
\includegraphics[width=0.95\textwidth]{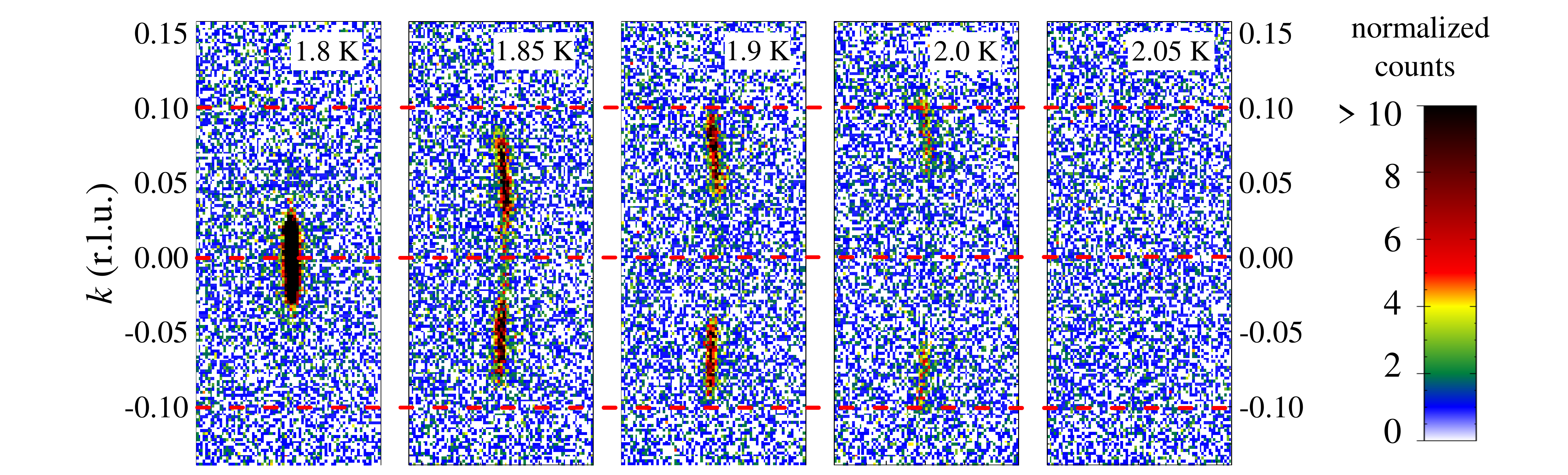}
\end{center}
\caption[1]{Detector images of the magnetic Bragg peaks recorded at E2 for selected temperatures in an external magnetic field of 6\,T $\parallel b$ axis. The direction out of the scattering plane was scaled as $k$ in units of r.l.u. The change in the appearance of the magnetic Bragg peaks represents the phase borderline IV--V. The dashed red lines indicate $k = 0$, $\pm 0.1$ r.l.u.}
\label{fig:EvolutionPeak6T}
\end{figure*}

\begin{figure}[b!]
\begin{center}
\includegraphics[width=0.9\columnwidth]{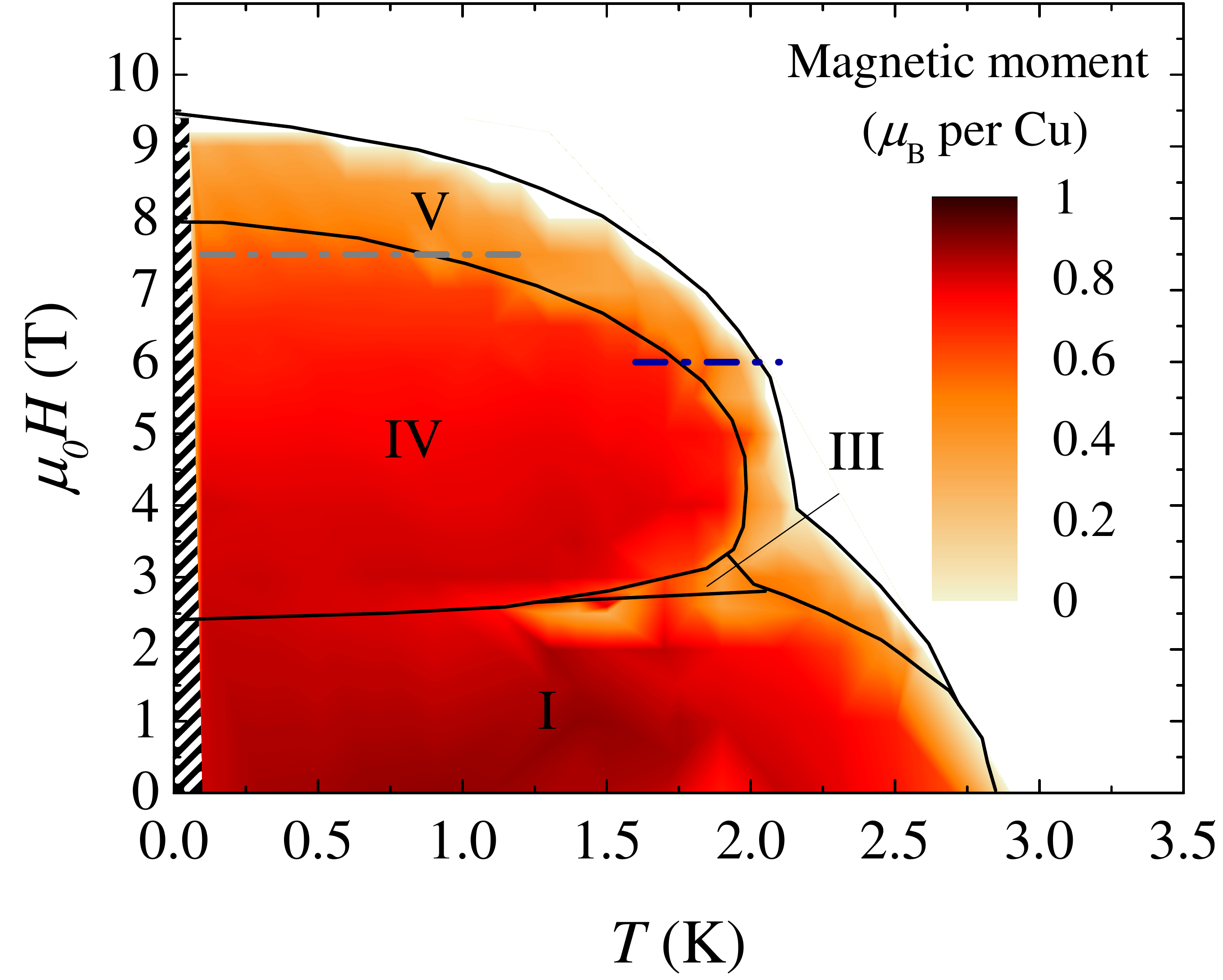}
\end{center}
\caption[1]{The contour plot of the magnetic moment $m$ per Cu atom for the different magnetic phases. On passing the phase boundaries I--V and IV--V there is an abrupt drop of the magnetic moment. The dashed grey and blue lines denote the scans from the Figs.~\ref{fig:Overview_kscans_D10_7p5T} and \ref{fig:EvolutionPeak6T}. In the black-and-white striped low temperature area no neutron diffraction scans were carried out. The phase borderlines were added to the contour plot by copying them from the phase diagram in Ref.~\cite{willenberg2016}.} 
\label{fig:MappingPhaseDiagram}
\end{figure}

The main focus of the present set of neutron diffraction experiments is the field and temperature dependence of the (integrated) neutron scattering intensity of the magnetic reflections present within the different magnetic phases/regions for $H \parallel b$ axis -- with special emphasis on the high-field phase V. As examples, in the Figs.~\ref{fig:Overview_kscans_D10_7p5T} and \ref{fig:EvolutionPeak6T} we present plots of the raw data from the D10 and E2 experiments. First, for the D10 study, in Fig.~\ref{fig:Overview_kscans_D10_7p5T} scans of (0 $k$ 1/2) with varying $k$ are presented for various temperatures between 0.1 and 1.2\,K in a magnetic field of 7.5\,T (the dashed gray line in Fig.~\ref{fig:MappingPhaseDiagram} denotes the scan in the magnetic phase diagram). For all examined temperature points up to 0.8\,K, a magnetic Bragg peak at (0 0 1/2) is observed, which corresponds to the commensurate antiferromagnetic spin alignment in phase IV, in accordance with the magnetic phase diagram established previously~\cite{willenberg2016}. The peak looses intensity abruptly at 0.9\,K and vanishes at 1.0\,K where two incommensurate magnetic Bragg peaks appear instead at (0 $\pm k_y$ 1/2) corresponding to the incommensurate longitudinal SDW in phase V~\cite{willenberg2016}.

In Fig.~\ref{fig:EvolutionPeak6T}, detector images of the magnetic reflections recorded at the instrument E2 are presented for selected temperatures between 1.80 and 2.05\,K in an external magnetic field of 6\,T (the dashed blue line in Fig.~\ref{fig:MappingPhaseDiagram} indicates the scan in the magnetic phase diagram). Here, the vertical axis of each image is scaled as $k$ axis. The scaling factor has been determined via the reference scan carried out in phase I, with its ordering vector $\vec{q} =$ (0 0.186 1/2)~\cite{willenberg2012}. At 1.80\,K, the commensurate magnetic Bragg peak at (0 0 1/2) is observed, which has vanished at 1.85\,K. Here, instead, two incommensurate magnetic Bragg peaks appear at (0 $\pm k_y$ 1/2). The peak intensity decreases towards 2.05\,K, at which point no further neutron scattering intensity is observed. 

Again, this change in the appearance of the magnetic Bragg peaks represents the phase borderline IV--V. The (dis)appearance of the magnetic Bragg peaks is consistent with the phase diagram shown in Fig.~\ref{fig:PhaseDiagram}. However, from Fig.~\ref{fig:EvolutionPeak6T}, a significant shift of $k_y$ with temperature is observed at 6\,T which is not present at 7.5\,T (see Fig.~\ref{fig:Overview_kscans_D10_7p5T}). This observation will be discussed in more detail later in this section.

For the determination of the peak position as well as the integrated intensity of the magnetic Bragg peaks the data were fitted using a Gaussian function (see Fig.~\ref{fig:Overview_kscans_D10_7p5T} for example fits to the scans performed at D10). From the integrated intensities $I$ of all magnetic Bragg peaks scanned the magnitude of the magnetic moment $m$ within the whole magnetic phase diagram for $H \parallel b$ axis was obtained. For an absolute scaling, the magnetic moment per Cu atom within the different phases had been determined before by a refinement of neutron diffraction data collected in phases I ($\mu_y = 0.833(10)\,\mu_{\rm B}$ and $\mu_{xz} = 0.638(15)\,\mu_{\rm B}$ at 1.8\,K, 0\,T), III, IV (0.79(1)\,$\mu_{\rm B}$ at 1.7\,K, 4\,T) and V (0.44(1)\,$\mu_{\rm B}$ at 1.9\,K, 6\,T)~\cite{willenberg2016, willenberg2012}. Since the magnetic moment $m^2 \sim I$, the square rooted integrated intensities were scaled linearly onto the magnetic moment values given above. For the data collected within the elliptical helix phase I, the magnetic moment along the $b$ axis, $\mu_y = 0.833(10)\,\mu_{\rm B}$, was used. Within the coexistence phase III, the commensurate magnetic Bragg peak at (0 0 1/2) was scanned.

The mapping of the magnetic moment per Cu atom is depicted in Fig.~\ref{fig:MappingPhaseDiagram} in form of a contour plot. Generally, within the low and intermediate magnetic field regions, the essential features of the magnetic phase diagram of linarite can be seen in the contour plot of the magnetic moment. Here, the phase boundaries, which were added to the contour plot in Fig.~\ref{fig:MappingPhaseDiagram} by copying them from the phase diagram in Ref.~\cite{willenberg2016}, separate the regions where neutron diffraction "sees" a different magnetic behavior. Especially, the phase transitions I--V and IV--V occur by an abrupt drop of the magnetic moment leading to a red coloring of the phase I and IV regions whereas the surrounding phase V is colored in orange representing a reduction in the moment by about a factor of two. As the scans were carried out by setting the magnetic field and then varying temperature stepwise, the region II, where magnetization hysteresis has been observed in previous measurements~\cite{willenberg2012}, is not seen in the neutron diffraction measurements at E2 and D10. 

\begin{figure}[b!]
\begin{center}
\includegraphics[width=0.9\columnwidth]{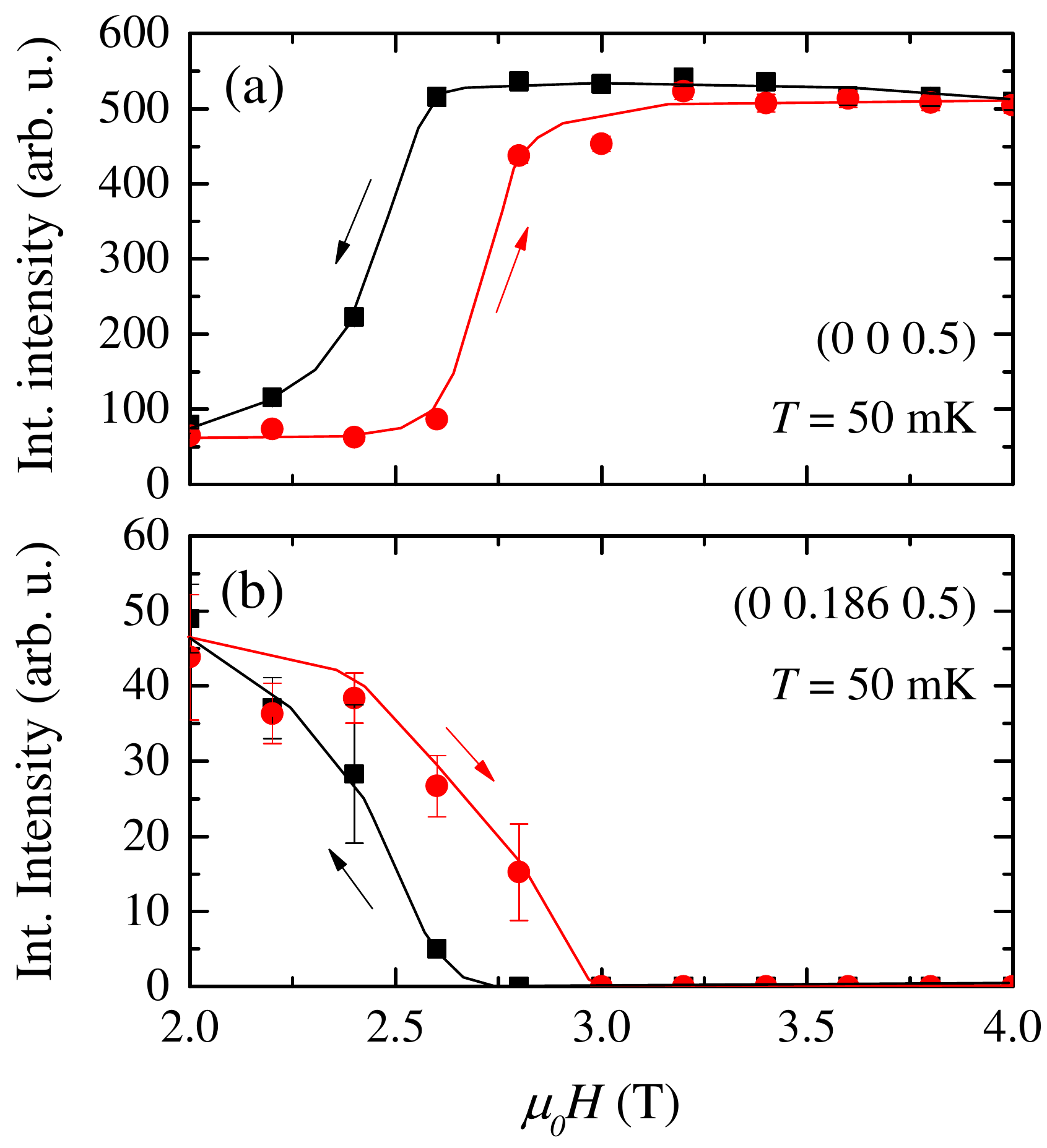}
\end{center}
\caption[1]{Hysteretic behavior of (a) the commensurate (0 0 1/2) peak and (b) the incommensurate (0 0.186 1/2) peak in region II at 50\,mK for increasing (red)/decreasing (black) magnetic field; solid lines are guides to the eye.} 
\label{fig:Hysteresis_WOMBAT}
\end{figure}

Instead, the measurements at WOMBAT were carried out by stabilizing the temperature and varying the applied field. These neutron diffraction data show that in region II is not a new magnetic phase but a coexistence of phase I and IV. Further, here, a hysteretic behavior of the incommensurate and commensurate magnetic Bragg peaks was observed, {\it i.e.}, at $T = 50$\,mK for applied fields from 2 to 3.5\,T and at $T = 300$\,mK from 2.25 to 3\,T. In Fig.~\ref{fig:Hysteresis_WOMBAT} (a) the field dependence of the commensurate magnetic (0 0 1/2) peak at 50\,mK is shown with increasing and decreasing magnetic field between 2 and 4\,T. Correspondingly, in Fig.~\ref{fig:Hysteresis_WOMBAT} (b) the field dependence is shown for the incommensurate (0 0.186 1/2) reflection. In contrast to phase III, the strong hysteresis observed in region II indicates a pinning of spins at this low temperature.

\begin{figure*}[t!]
\begin{center}
\includegraphics[width=0.95\textwidth]{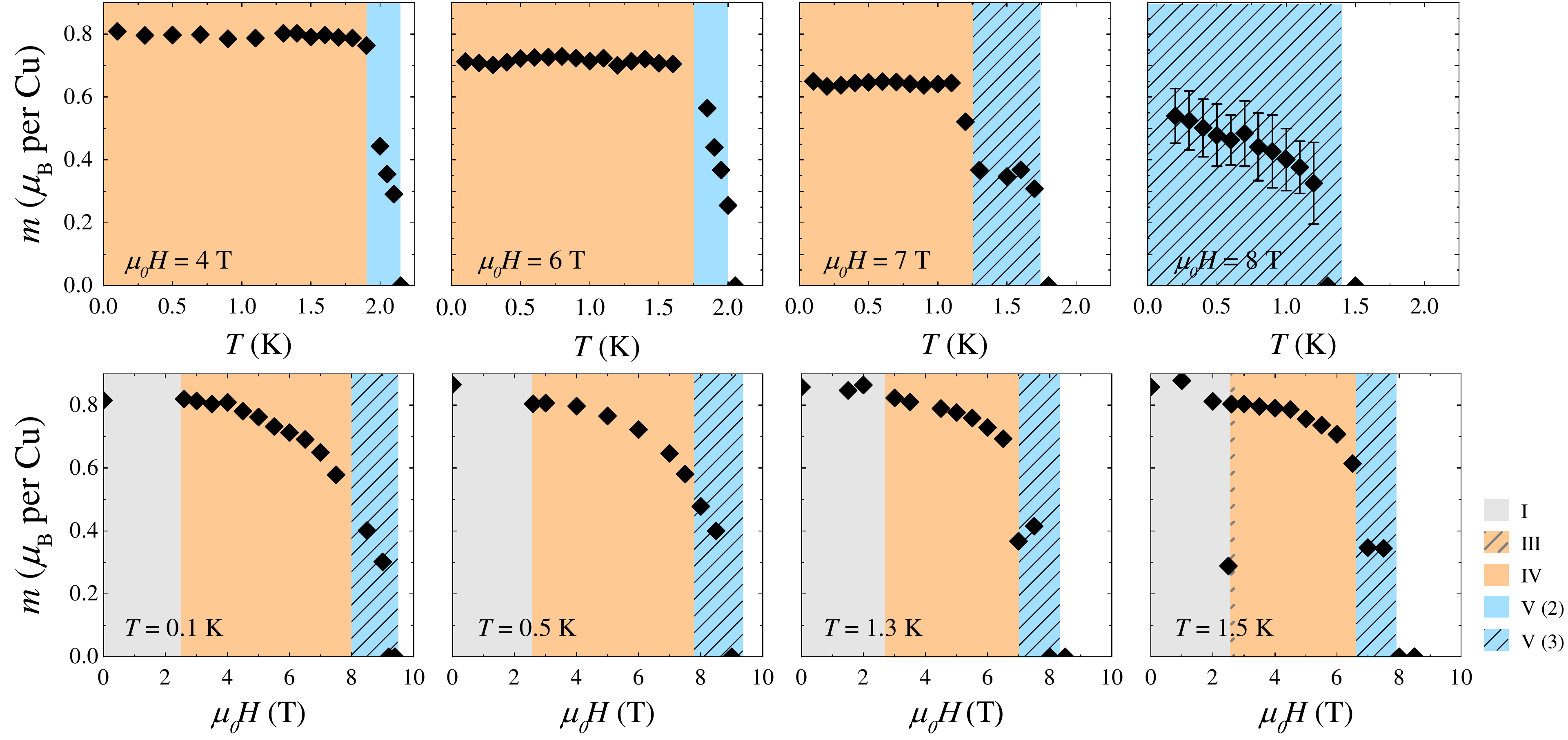}
\end{center}
\caption{First row: Cuts through the magnetic moment map in Fig.~\ref{fig:MappingPhaseDiagram} at various fields. At the phase boundary IV--V, a step-like decrease of the magnetic moment is observed for all magnetic fields. The background color in each plot corresponds to the coloring of the magnetic phases in Fig.~\ref{fig:PhaseDiagram}. The notation "phase V (1), (2), (3)" is defined later in the text and described in Fig.~\ref{fig:PhaseDiagram}. The phase boundaries were drawn according to the phase diagram from the thermodynamic measurements on linarite~\cite{schaepers2013,willenberg2012,willenberg2016}. In the diagram for the 8\,T data, the error bars for the magnetic moment values in phase V represent the typical error size for the phase V magnetic moments. The magnetic moment error in phase I and IV is of the order of the data point size. Second row: Cuts through magnetic moment map at various temperatures. Note that the cut at 1.5\,K passes through phase III where both reflections at (0 0 1/2) and (0 0.186 1/2) are present.}
\label{fig:Linarite_Cuts_PhaseDiagram_T}
\end{figure*}

To further illustrate the field and temperature behavior of the magnetic moment $m$, cuts through the contour plot of the magnetic moment in Fig.~\ref{fig:MappingPhaseDiagram} along the temperature and the field axis are presented in Fig.~\ref{fig:Linarite_Cuts_PhaseDiagram_T} for four selected magnetic fields (4, 6, 7, and 8\,T) and four selected temperatures (0.1, 0.5, 1.3 and 1.5\,K). Within phase IV, for all magnetic fields studied, the magnetic moment, {\it i.e.}, the intensity of the magnetic Bragg peaks, stays almost constant when increasing temperature. However, when passing the phase boundary IV--V, there is always a sudden drop of the magnetic moment resulting in a step-like appearance of the magnetic moment when varying $T$ (with a maximum magnetic moment change by a factor of $\sim 2$). This, in return, means that the magnetic moment does not change continuously at the phase boundary IV--V, which indicates that this phase transition is of first order nature. The cut along $T$ at 8\,T is completely within phase V.

When cutting the magnetic moment map along the field axis, $m(\mu_0 H)$ continuously decreases with field, and somewhat more slowly in the low magnetic field region of phase IV ($\mu_0 H < 4$\,T) than in the high field region ($\mu_0 H > 4$\,T). During a previous neutron diffraction study on phase IV, also a decrease of the antiferromagnetic magnetic moment with increasing magnetic field has been observed, together with additional scattering intensity on top of nuclear Bragg peaks. This indicated that the spins are driven into field-induced polarization~\cite{willenberg2012}. From phase IV into phase V, again a step-like decrease is observed, especially for the cuts at 1.3 and 1.5\,K, consistent with the phase transition IV--V being of first order.

In addition to the magnitude of the magnetic moment, the position of the incommensurate magnetic Bragg peak at (0 $k_y$ 1/2) in phase V was studied. In a previous neutron diffraction experiment it has been observed that $k_y$ shifts significantly when varying the magnetic field~\cite{willenberg2016}. Only, the shift is clearly different from theoretical predictions for the isotropic frustrated $J_1$-$J_2$ chain in Ref.~\cite{sudan2009}. Starting from these early observations, from the present neutron diffraction experiment, we have set out to fully establish the field and temperature dependence of $k_y$ in phase V. In order to correct for a possible experimental offset in the $k_y$ determination, the absolute values for the two incommensurate Bragg peaks at $\pm k_y$ obtained from the Gaussian fits to the D10 neutron diffraction data (see Fig.~\ref{fig:Overview_kscans_D10_7p5T}) were averaged. For the E2 data, the reference point recorded in phase I was used for the determination of the $k_y$ value from the peak center.

\begin{figure}[t!]
\begin{center}
\includegraphics[width=\columnwidth]{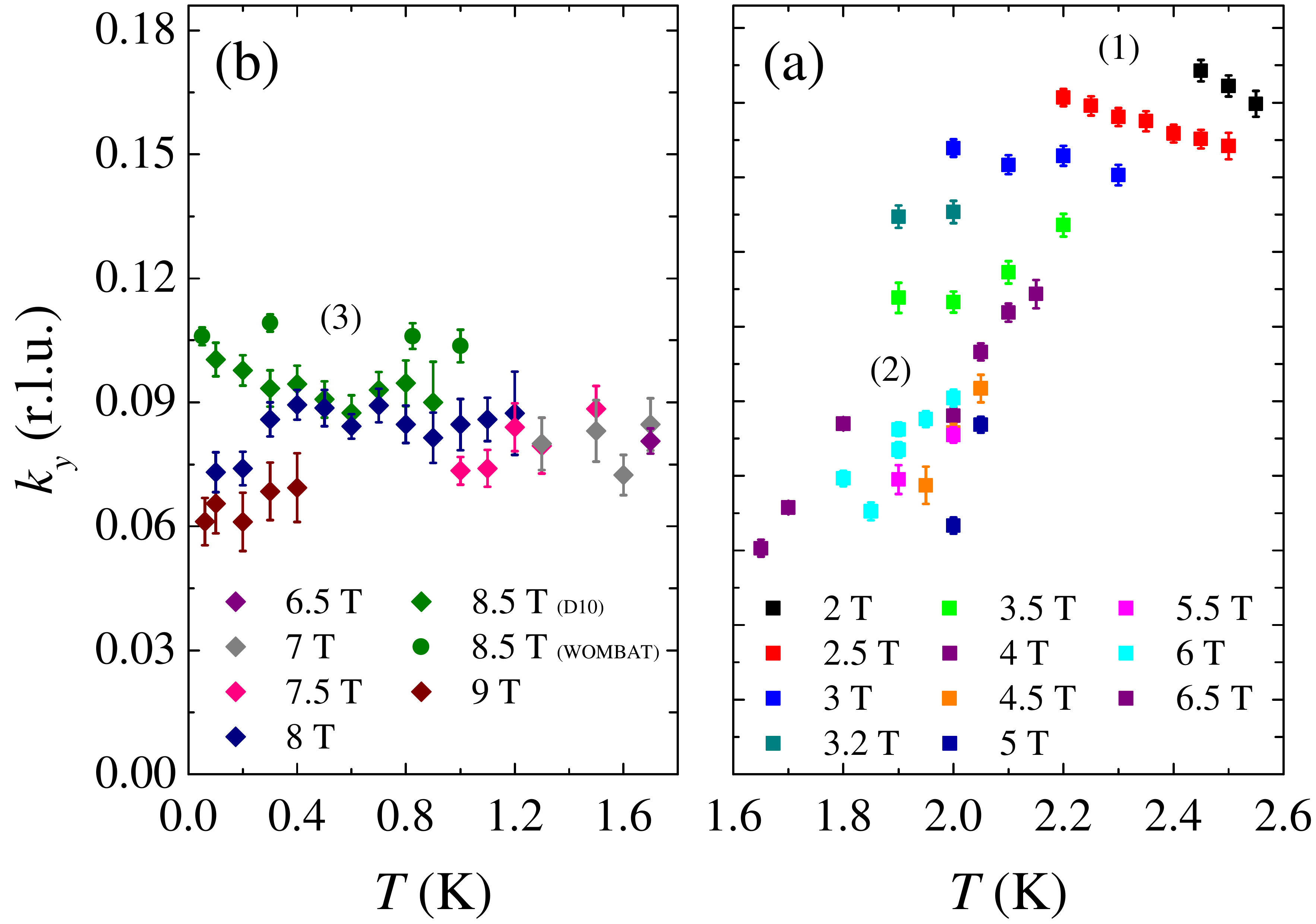}
\end{center}
\caption[1]{Temperature dependence of the SDW propagation vector component $k_y$ in magnetic fields $H \parallel b$ axis. The temperature behavior of $k_y$ separates phase V into three different regions: (1) the low field region where $k_y$ decreases when increasing $T$, (2) the intermediate field region with $k_y$ increasing when increasing $T$ and (3) the high field region where there is no significant $T$ dependence of $k_y$.} 
\label{fig:PropagationVector}
\end{figure}

From our analysis, we observe an even more complex behavior of $k_y (T, \mu_0 H)$ where $k_y$ does not only show a field but also temperature dependence (see Fig.~\ref{fig:PropagationVector}). The temperature dependence of $k_y$ basically separates the phase V into three regions (see Fig.~\ref{fig:PhaseDiagram}): (1) The low magnetic field region ($\mu_0 H < 3.2$\,T) where $k_y$ decreases when increasing temperature, (2) the intermediate field region (3.2 to 6.5\,T) where $k_y$ increases when increasing temperature and (3) the high magnetic field region ($\mu_0 H > 6.5$\,T) where no significant change of $k_y (T)$ with temperature is observed. The magnetic field $\mu_0 H \sim 3.2\,T$ acts as a "turning point" between (1) and (2), with $k_y$ essentially being temperature independent.

Qualitatively, the behavior of $k_y (T)$ in phase V relates to the adjoining ordered phase within the respective magnetic field region: In low fields, the incommensurability vector component $k_y$ in phase V increases as temperatures decreases towards the incommensurability of the helical phase with a magnetic propagation vector (0 0.186 1/2). In intermediate fields, for decreasing temperature the incommensurability $k_y$ closes in onto the commensurate state in phase IV. Finally, in the high-field region, the incommensurability of $k_y$ appears to be independent from adjoining phases. We stress that this evolution of $k_y$ is much more complex than stated in Ref.~\cite{Cemal2018} where only a single field dependence of $k_y$ at 60\,mK (and without reporting the experimental error) was given.

\subsection{Nuclear magnetic resonance (NMR)}

\begin{figure}[b!]
\begin{center}
\includegraphics[width=0.9\columnwidth]{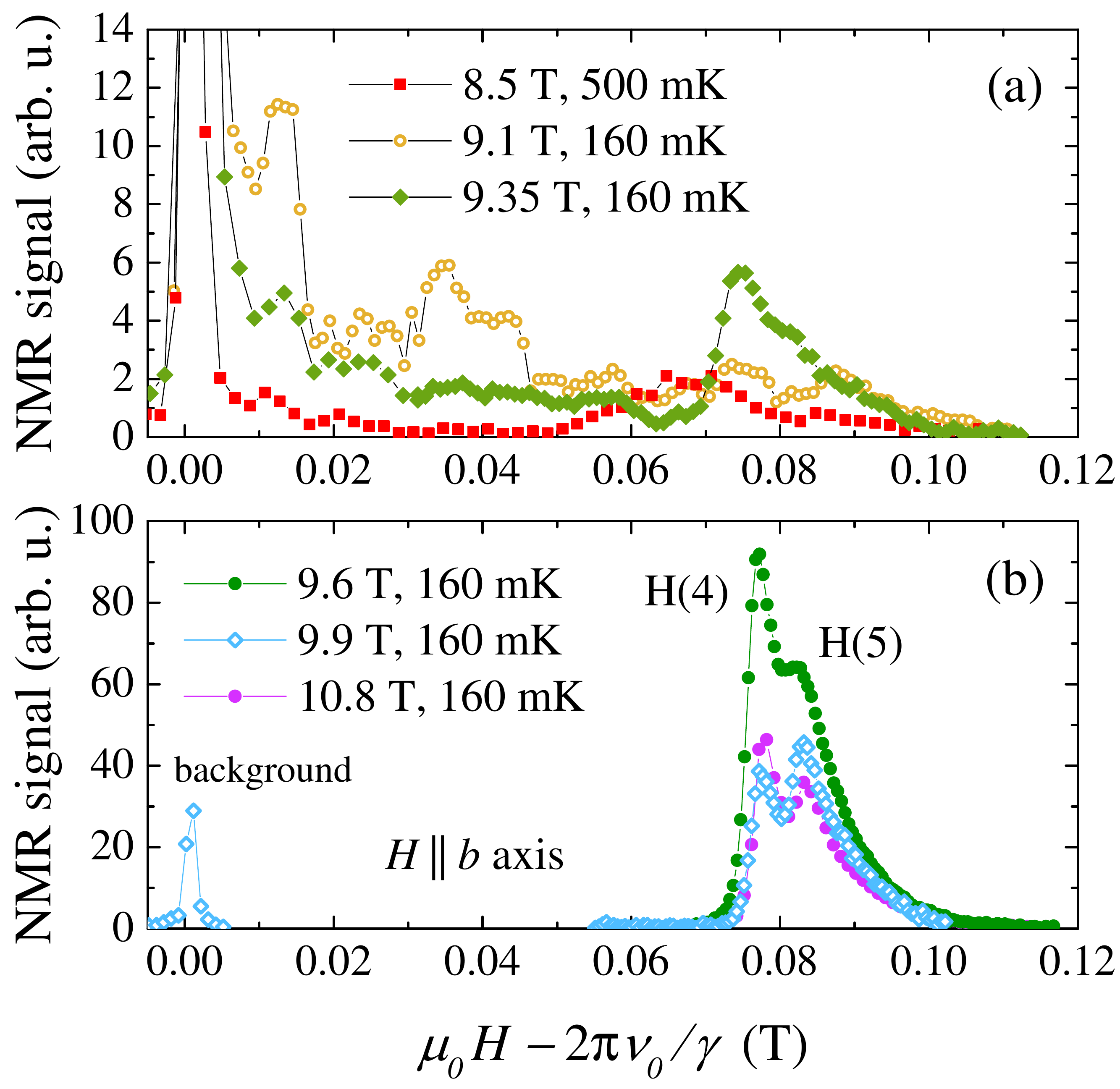}
\caption{(a) NMR spectra for different magnetic fields applied along the $b$ axis as a function of the relative field $\mu_0 H - 2 \pi \nu_0 / \gamma$. (b) NMR spectra for fields at 9.6\,T and above represented on a different scale. The background signal around $\mu_0 H - 2 \pi \nu_0 / \gamma = 0$\,T is expected to vary slowly with field and was determined separately only for $\mu_0 H = 9.9$\,T. The two other peaks are associated with the hydrogen sites H(4) and H(5)~\cite{Schaepers2014}.} 
\label{NMRspectrum}
\end{center}
\end{figure}

The low temperature NMR spectra are presented in Fig.~\ref{NMRspectrum} (a) and (b) for different magnetic fields applied along the $b$ axis. They were measured as a function of the magnetic field at a constant frequency $\nu_0$ and are plotted as function of the relative field $\mu_0 H - 2 \pi \nu_0 / \gamma$, where $\gamma$ represents the gyromagnetic ratio and $\nu_0$ is the frequency of the oscillating field. For all data sets there is a NMR signal at zero frequency shift stemming from the experimental set-up.

At $\mu_0 H = 8.5$\,T and $T = 500$\,mK, in phase V, in addition to the zero shift signal the NMR spectrum shows a broad peak resulting from the sample at $\mu_0 H - 2 \pi \nu_0 / \gamma = 0.068$\,T. Previous measurements in phase V at lower magnetic field and higher temperature ($\mu_0 H \leq 7$\,T and $T \geq 1.7$\,K) showed a superposition of the SDW signal and a broad peak similar to a paramagnetic signal, which becomes more pronounced with increasing magnetic field~\cite{willenberg2016}. The measurement reported here at a higher magnetic field of $\mu_0 H = 8.5$\,T thus probably only shows the broad peak corresponding to the non-dipolar ordered part of the sample. This peak with a full width at half maximum (FWHM) of about 200\,Oe is broader than in the previous measurements at $T = 1.7$\,K and $\mu_0 H = 7.5$\,T, which showed a FWHM of 0.4\,MHz corresponding to 100\,Oe. The signature of the SDW appears to be smeared/wiped out. This might arise from a small $T_2$ or a diminution of the volume fraction of the SDW state.

Next, at $\mu_0 H = 9.1$\,T and $T = 160$\,mK, a broad NMR signal is observed with a finite intensity from zero to 0.11T NMR shift. This spectrum does not show a clear peak resembling a paramagnetic signature. Magnetization, magnetocaloric effect and polarization current measurements do not show any signature of this change~\cite{willenberg2012,schaepers2013,Mack2017}.

At $\mu_0 H = 9.35$\,T and $T = 160$\,mK, in addition to the broad NMR signal an asymmetric peak is observed at a shift of 0.075\,T. This peak becomes the dominant feature, with an increase of the NMR signal by more than an order of magnitude, as the field is further increased to 9.6\,T. The strong increase together with a constant shift (within our resolution) and a decrease of the line width indicates the quasi-saturation of the magnetization. The NMR peak at $\mu_0 H = 9.6$\,T shows a shoulder and evolves into a double peak at $\mu_0 H = 9.9$\,T, which is unchanged at least up to $\mu_0H=10.8$~T again supporting the notion of magnetic quasi-saturation. The two-peak structure stems from the two inequivalent hydrogen sites H(4) and H(5) with slightly different hyperfine couplings for $H \parallel b$ axis, as previously discussed from measurements at higher temperature~\cite{wolter2012,Schaepers2014}.

\begin{figure}[b!]
\begin{center}
\includegraphics[width=\linewidth]{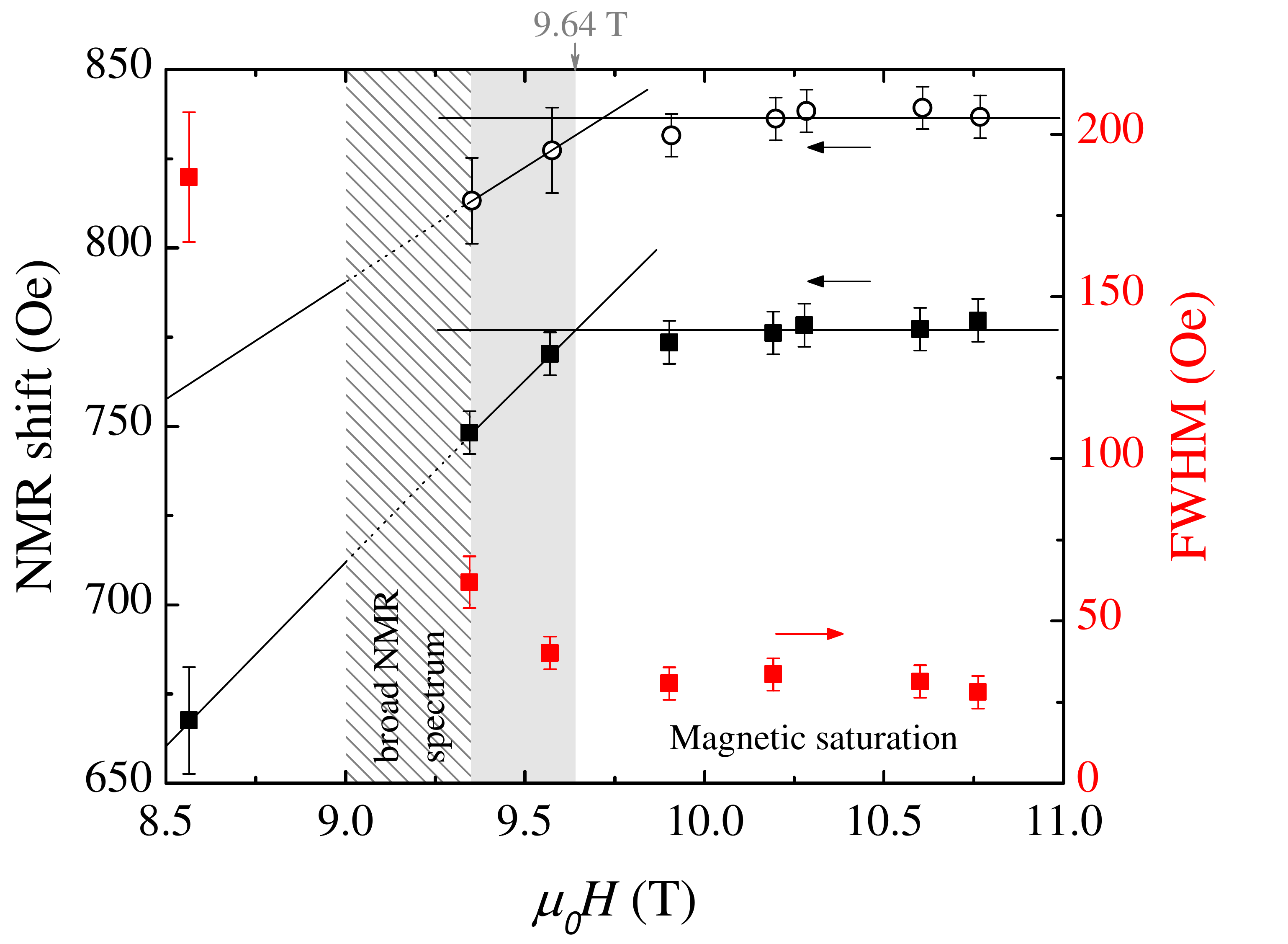}
\caption{NMR shift and FWHM as a function of the applied magnetic field $H \parallel b$ axis. The full and open symbols represent the H(4) and H(5) sites, respectively. The lines are guides to the eye and indicate the (quasi-)saturation field at $\mu_0 H_{\rm sat} = 9.64 \pm 0.10$\,T. The light grey area indicates the field interval where the occurrence of a multipolar state is possible. At 9.1\,T the broad NMR spectrum has been observed. The measurements were carried out at 160\,mK except for the 8.5\,T data which were taken at 500\,mK and serve as reference.} 
\label{NMRshift}
\end{center}
\end{figure}

The NMR shift for both hydrogen sites and the FWHM for the hydrogen site H(4) at $T = 160$\,mK are represented as a function of field in Fig.~\ref{NMRshift} for $H \parallel b$ axis. This microscopic measurement at a low temperature of 160\,mK enables us to finally determine the quasi-saturation field with a much higher accuracy than previous magnetization measurements performed at 1.8\,K~\cite{wolter2012}. The observation of a constant NMR shift together with a narrow NMR line above $\mu_0 H_{\rm sat} = 9.64 \pm 0.10$\,T is indicative of the quasi-saturation field. In the field interval 9.35\,T\,$<$\,$\mu_0 H$\,$<$\,9.64\,T, the absence of the SDW signature in both NMR and neutrons together with a narrow NMR line plus a non-constant NMR shift indicates the possible realization of a multipolar state in this field interval as it was proposed by similar NMR studies on the $J_1$-$J_2$ Heisenberg chain systems LiCuVO$_4$~\cite{Orlova2017} and LiCuSbO$_4$~\cite{Bosiocic2017}. This result is in contradiction with the theoretical proposal of a direct phase transition from phase V into magnetic saturation~\cite{Cemal2018} and with the interpretation of recent torque magnetometry measurements localizing this transition at $\mu_0 H = 9.3$\,T for $T = 0.2$\,K~\cite{Feng2018}.

The study of the NMR spin lattice relaxation rate in the region close to magnetic (quasi-)saturation allows us to probe the existence and nature of bound magnons, which could be responsible for the formation of a multipolar state. Indeed, at the magnetic saturation, a spin gap $\Delta$ opens and this gap becomes larger when the field is increased further above the saturation field. In the particular case of a $J_1$-$J_2$ Heisenberg chain, the slope $d\Delta / d \mu_0 H$ is given by $d\Delta / d \mu_0 H = p g \mu_{\rm B} / \mu_0 k_{\rm B}$ where $p$ is an integer number such that $p$-bound magnon excitations are the lowest energy excitations in the saturated regime~\cite{zhitomirskyLiCuVO4}. Thus, the study of the field dependence of the spin gap at the magnetic saturation from NMR spin relaxation rate measurements allows us to identify and characterize the bound magnons. 
 
In addition, the study of NMR spin relaxation rates below the magnetic saturation in a $J_1$-$J_2$ Heisenberg chain was proposed as a way to identify the multipolar state~\cite{sato2009,sato2011,smerald2016}. Taking into account the monoclinic symmetry, the relaxation rate in a magnetic field along the $b$ axis $(1/T_1)_b$ can be expressed in an orthorhombic coordinate system ${a_\perp, b, c}$~\cite{Goto2006,nawa2013LiCuVO4} as
\begin{eqnarray}
(1/T_1)_b = && \frac{\gamma^2}{2N}\sum_q C_{a_\perp}(q) S_{a_\perp a_\perp}(q,\omega) + C_b(q) S_{bb}(q,\omega)\nonumber\\
&& + C_c(q)S_{cc}(q,\omega) + C_{a_\perp c}(q)S_{a_\perp c}(q,\omega),
\end{eqnarray}
where $\gamma$ and $N$ stand for the gyromagnetic ratio and the number of Cu atoms in the system respectively. $C_{a_\perp}(q)$, $C_b(q)$, $C_c(q)$ and $C_{a_\perp c}(q)$ are the geometrical form factors in momentum space. The dynamical spin correlation function $S_{\mu\nu}(q,\omega)$ for the crystallographic axis $\mu$ and $\nu$ is defined as
\begin{eqnarray}
S_{\mu\nu}(q,\omega) = \int_{-\infty}^{+\infty} & dt\,e^{i\omega t}(S_\mu(q,t) S_\nu(-q,0)\nonumber\\
& + S_\nu(-q,0)S_\mu(q,t))/2.
\end{eqnarray}
Here, $S(q,t)$ is the Fourier transform of the surrounding electron spins $S(r_i,t)$ located at the Cu site $r_i$ with
\begin{equation}
S(q,t) = \frac{1}{\sqrt{N}} \sum_i S(r_i,t) e^{-i q_i r_i}.
\end{equation}
While the transverse dynamical spin correlation $S_{a_\perp a_\perp}$, $S_{cc}$ and $S_{a_\perp c}$ were predicted to show an exponential spin gap in the bound magnon state~\cite{vekua2007,zhitomirskyLiCuVO4,Syromyatnikov2012,starykh2014} the longitudinal dynamical spin correlation $S_{bb}$ was predicted to follow a power law behavior in temperature~\cite{Chitra1997,sato2009,sato2011}. The NMR relaxation rate was also studied theoretically in the low-temperature limit and a step at the multipolar ordering from the longitudinal spin fluctuations was predicted~\cite{smerald2016}. However, the contribution from longitudinal spin fluctuations to the spin-lattice relaxation rate is proportional to the form factor
\begin{equation}
C_b(q)=(|A_{a_\perp b}(q)|^2+|A_{bc}(q)|^2)g_{bb}^2,
\end{equation}
where $A(q)$ and $g$ stand for the hyperfine coupling tensor in Fourier space and the $g$ tensor, respectively. In linarite, for $H \parallel b$ axis, the hyperfine coupling coefficients $A_{a_\perp b}$ and $A_{bc}$ are suppressed by the symmetry of the crystal structure~\cite{Schaepers2014} and, as a consequence, the form factor $C_b(q)$ vanishes. In this particular case, the relaxation rate in the multipolar state depends only on the gapped transverse fluctuations and must follow the Arrhenius law
\begin{equation}
(1/T_1)_b \propto e^{-\Delta/T}, \label{eq_gap}
\end{equation}
where $\Delta$ represents the spin gap. This gap would arise from the condensation of bound magnons and harbor single magnons as lowest energy excitations~\cite{vekua2007}. As a consequence, the field dependence of the spin gap would be less steep than in the (quasi-)saturated regime $d\Delta/d \mu_0 H = g \mu_{\rm B} / \mu_0 k_{\rm B}$ = 1.41\,K/T, using a $g$ factor $g_b$ = 2.10 obtained by previous ESR experiments on linarite~\cite{wolter2012}.

\begin{figure}[t!]
\begin{center}
\includegraphics[width=0.95\linewidth]{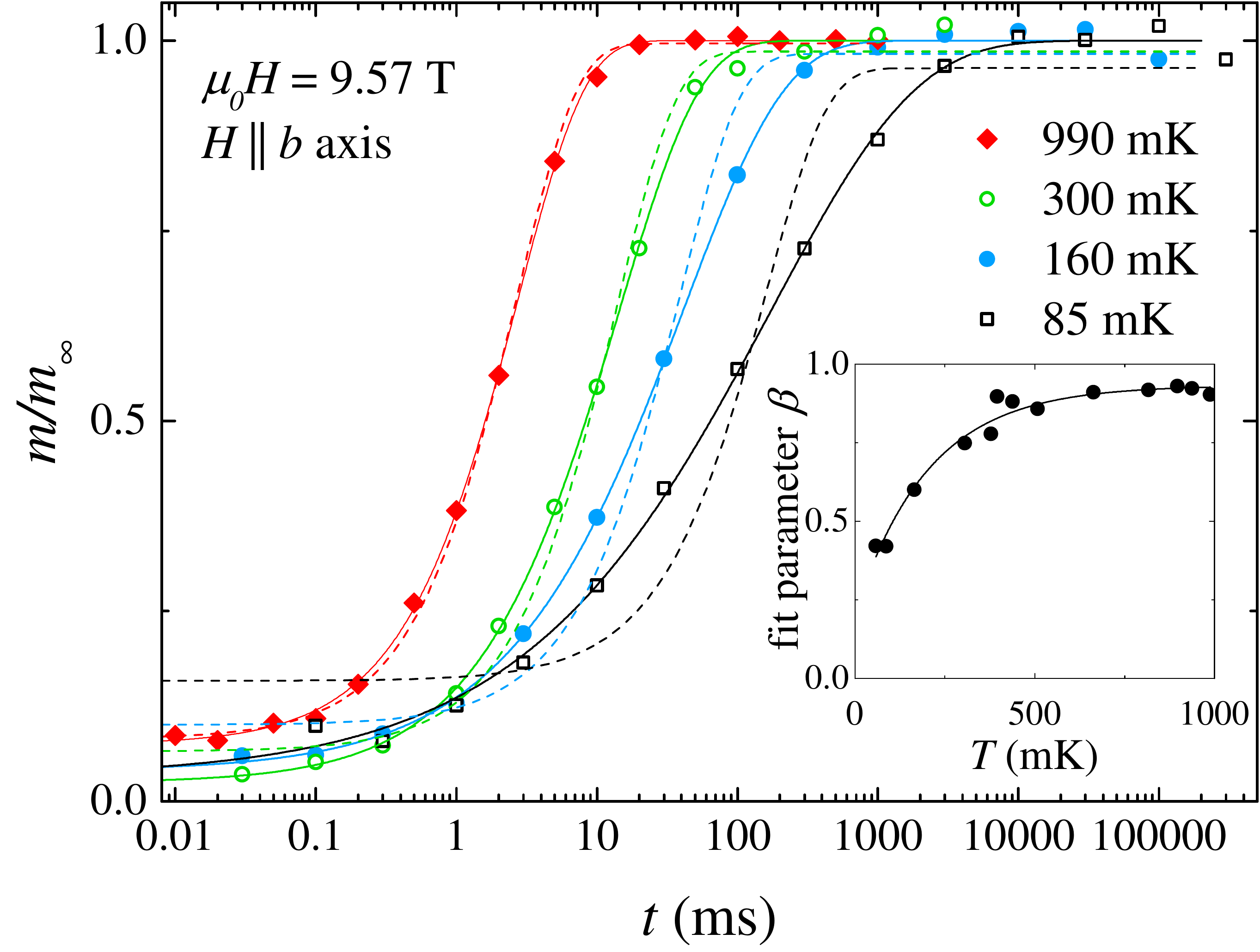}
\caption{Normalized magnetization recovery $m/m_\infty$ of the hydrogen nuclei H(4) as function of time $t$ after its suppression by $\pi/2$ pulses at $\mu_0 H = 9.57$\,T and different temperatures. The dashed and solid lines are fits to Eqs.~(\ref{eq_T1}) and (\ref{eq_T1beta}), respectively. The inset shows the fit parameter $\beta$ as a function of temperature. The solid line in the inset is a guide to the eye.} 
\label{T1m}
\end{center}
\end{figure}

The field and temperature dependence of the NMR spin-lattice relaxation rate $1/T_1$ was measured at the summit of the peak of H(4) to be compared with these predictions. The nuclear magnetization $m$ at this hydrogen site is represented in Fig.~\ref{T1m} as a function of time $t$ after the saturation $\pi/2$ pulses for $\mu_0 H = 9.57$\,T for different temperatures. Down to $T \sim 300$\,mK, it can be well fitted according to the equation
\begin{equation}
m = m_\infty (1 - f e^{-t/T_1}), \label{eq_T1}
\end{equation}
where $m_\infty$ is the value in equilibrium state and $f$ is a fit parameter to account for a non-perfect suppression of the nuclear magnetization of this hydrogen site. Below $T \sim 300$\,mK, however, a clear deviation from this behavior is observed indicating a distribution of $T_1$ relaxation times. Therefore, the magnetization of the hydrogen nuclei $m$ was fitted according to
\begin{equation}
m = m_\infty (1 - f e^{(-t/T_1)^\beta}), \label{eq_T1beta}
\end{equation}
where the stretching exponent $\beta < 1$ is related to the distribution of $T_1$ times~\cite{Johnston2006}. Using Eq.~(\ref{eq_T1beta}) it was possible to fit the curves in Fig.~\ref{T1m} down to the lowest temperatures. The best fit parameter $\beta$ is included as a function of temperature in the inset of Fig.~\ref{T1m}. It increases with temperature and saturates at a value close to the one at $T \sim 300$\,mK. The distribution of $T_1$ relaxation times at low temperatures may arise from intrinsic local excitations as it was already proposed for another $J_1$-$J_2$ Heisenberg chain system LiCuSbO$_4$, although in linarite this behavior occurs at a temperature one order of magnitude lower than in the latter system~\cite{grafe2017,Bosiocic2017}.

\begin{figure}[t!]
\begin{center}
\includegraphics[width=0.9\linewidth]{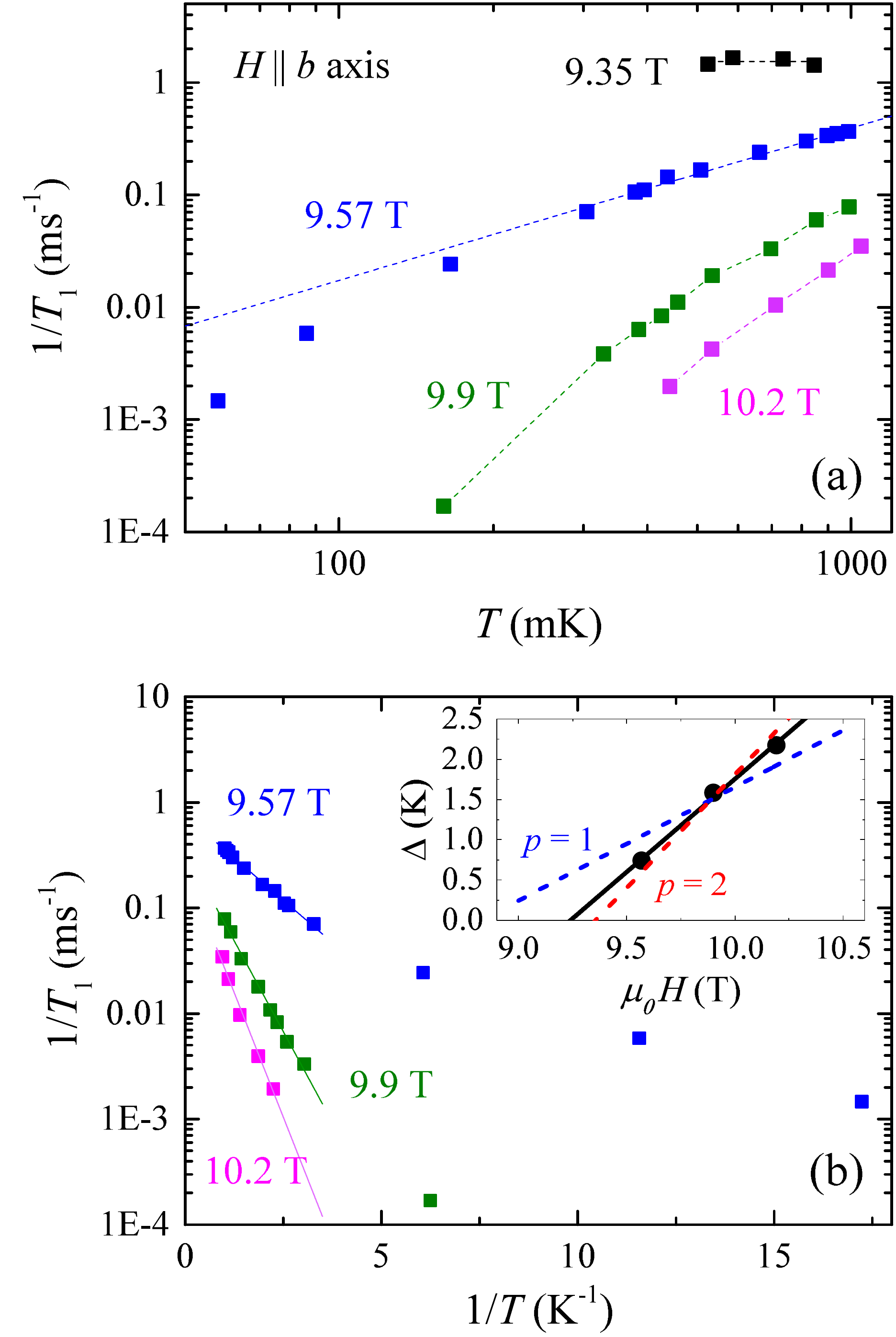}
\caption{(a) Spin-lattice relaxation rate $1/T_1$ as a function of temperature $T$ and magnetic field $H \parallel b$ axis. The lines are guides to the eye. (b) Relaxation rate $1/T_1$ as a function of the inverse temperature $1/T$. The lines are linear fits to Eq.~(\ref{eq_gap}) in the temperature range 300\,mK\,$<$\,$T$\,$<$\,1\,K. The inset shows the field dependence of the gap $\Delta$ extracted from the fits. The blue and red dashed lines correspond to the expected behavior for a one-magnon gap ($p = 1$) and a bound two-magnon gap ($p = 2$), respectively. The solid black line in the inset is a linear fit to our experimental data and gives $d\Delta / d \mu_0 H = (2.6 \pm 0.5)$\,K/T.} 
\label{T1}
\end{center}
\end{figure}

The resulting relaxation rate $1/T_1$ is represented as a function of temperature in Fig.~\ref{T1}(a) and of the inverse temperature $1/T$, together with fits to Eq.~(\ref{eq_gap}) in Fig.~\ref{T1}(b). At $\mu_0 H$ = 9.35\,T and for temperatures above the magnetic transition $T_{\rm V}(\mu_0 H$ = 9.35\,T)$\,\sim\,$400\,mK, $1/T_1$ is constant in temperature up to 1\,K. At $\mu_0 H = 9.57$\,T, $1/T_1$ increases with temperature over two orders of magnitude between 60\,mK and 1\,K. This temperature dependence cannot be fitted by Eq.~(\ref{eq_gap}) within the whole temperature range but it can be fitted by Eq.~(\ref{eq_gap}) for $T \geq 300$\,mK, where no broad distribution of $T_1$ is observed \cite{note1}. At $\mu_0 H = 9.9$\,T and $\mu_0 H = 10.19$\,T, {\it i.e.}, above the (quasi-)saturation field, where a full polarization of the electronic spins is already expected from the study of the NMR shift, the temperature dependence of the relaxation rate $1/T_1$ can also be well fitted by Eq.~(\ref{eq_gap}) down to 300\,mK.

The extracted gap values are represented in the inset of Fig.~\ref{T1} (b). Despite our limited statistics, one can see a clear increase of the gap values as function of the field following a linear behavior with a slope of $d\Delta / d \mu_0 H = (2.6 \pm 0.5)$\,K/T. The expected slopes for one- and two-magnon excitations, {\it i.e.}, 1.41\,K/T and 2.82\,K/T, are represented in the inset of Fig.~\ref{T1} (b) together with the experimentally extracted values. The experimental slope $d\Delta / d \mu_0 H = (2.6 \pm 0.5)$\,K/T is much closer to the two-magnon value, indicating that two-bound magnon excitations are indeed the lowest energy excitations. Actually, while the occurrence of three-bound magnons was first proposed for the frustration ratio $\alpha = J_2 / J_1 \sim -0.33$ of linarite~\cite{sudan2009} the predominance of two bound-magnon in the vicinity of the magnetic saturation was theoretically predicted as a consequence of the $XYZ$ exchange anisotropy~\cite{willenberg2016}. As a consequence, the spin multipolar state of linarite, if it exists, would be a spin quadrupolar (nematic) state. It should be emphasized that the two-bound magnon gap seems to open around $\mu_0 H \simeq 9.35$\,T, {\it i.e.}, already below the magnetic (quasi-)saturation ($\mu_0 H_{\rm sat} = 9.64$\,T), but where no dipolar magnetic order is observed.
 
We note that the exact meaning and the nature of these magnetic excitations remains unclear due to the non-conservation of $S_z$ in low-symmetry systems, to which linarite belongs ({\it i.e.}, $XYZ$ anisotropy and DMI). In this context one-, two- etc. magnons might be an approximate description, only. This consideration might be helpful to interpret the somewhat anomalous $g$-dependence or the missing one-magnon spin-gap especially if a dipolar component derived from fan-states and/or a special coexisting hypothetical spiral-like incommensurate new phase as proposed in Ref.~\cite{smerald2016} for BaCdVO(PO$_4$)$_2$ would be present in the nematic state. 

Overall, our findings show strong similarities to previous results on another $J_1$-$J_2$ Heisenberg chain compound, that is LiCuVO$_4$ in a magnetic field along the $c$ axis. Indeed, while the realization of a multipolar state was proposed in the field interval 42.41\,T $< \mu_0 H <$\,43.55\,T in LiCuVO$_4$ from the study of the NMR spectrum~\cite{Orlova2017}, a previous NMR relaxation rate study shows a gap opening for $\mu_0 H \geq 41$\,T with a slope corresponding to a two-magnon gap~\cite{buettgen2014}. Thus, this feature could be a rather general behavior of $J_1$-$J_2$ Heisenberg chains. In contrast, the observation of a single magnon gap was reported in another $J_1$-$J_2$ Heisenberg chain system, LiCuSbO$_4$, between 13\,T and 16\,T by NMR measurement on a powder sample and proposed as a signature of a multipolar state~\cite{grafe2017}. However, NMR measurements on oriented powder~\cite{Bosiocic2017} with the magnetic field along the hard magnetization axis showed the magnetic saturation at 13\,T. Thus the single magnon gap observed in Ref.~\cite{grafe2017} must be an average gap of the different field directions including directions in the saturated regime and NMR measurements on single crystals or oriented powder of LiCuSbO$_4$ would be valuable to confirm the observation of a single magnon gap.

\subsection{Theoretical aspects}

It is well known that the peculiarities of the crystal structure, especially the Cu--O--Cu bond angle in edge-sharing CuO$_2$ chains, are of crucial relevance for the size of the exchange integrals. Several studies~\cite{azurite,clinoclase} have shown that also the exact H-positions of the O--H ligands of Cu are of great importance for the strength of the magnetic couplings. Therefore, we have re-investigated the influence of the H-positions in linarite on the electronic structure and the main exchange interaction regarding the new single crystal neutron refinements and optimization of its two H-positions applying DFT calculations.

Using the structural parameters of Sch\"apers {\it et al.}~\cite{Schaepers2014} and fixing the internal coordinates of all positions except the H atoms (the heavier atoms are well determined from previous XRD and neutron diffraction experiments~\cite{effenberger,schaepers2013}), we find almost perfect agreement between the experimental and the calculated H-positions (H$_{\rm exp}$(4): [0.8667(4), 1/4, 0.6166(8)] vs. H$_{\rm calc}$(4): [0.8646, 1/4, 0.6169]; H$_{\rm exp}$(5): [0.0586(4), 1/4, 0.4537(7)] vs. H$_{\rm calc}$(5): [0.0548, 1/4, 0.4535]). The reliability of the DFT optimized H-positions has already been demonstrated studying the related Cu$^{2+}$ mineral malachite\cite{malachite}, with the most precise atomic coordinates resulting from the general gradient approximation for the exchange-correlation potential as applied here. The deviations of the O--H bond length between the neutron refinement and the DFT procedure are less than 0.02 \AA , the Cu--O--H bond angle differs by less than 3$^\circ$. 

Using these new structural parameters, the DFT derived values (applying a typical Coulomb repulsion $U_d = 7$\,eV) for the exchange parameters yield $J_1 = -121$\,K, $J_2 = 38$\,K and $J_2/J_1 = 0.31$, differing only slightly from the previously published DFT values\cite{wolter2012} ($J_1 = -133$\,K, $J_2 = 42$\,K and $J_2/J_1 = 0.32$. Considering the error bars of the calculational procedure, given for instance the unknown exact value of $U_d$ in the DFT+$U$ approach~\cite{Schaepers2014}, the crystal structure related aspect for the size of the exchange integrals in linarite can now be considered as fully settled. Of course, beyond the equilibrium H-position, its quantum fluctuations and thermal fluctuations at higher temperatures could be still of importance and should be studied in future investigations.

\begin{figure}[t!]
\begin{center}
\includegraphics[width=0.95\linewidth]{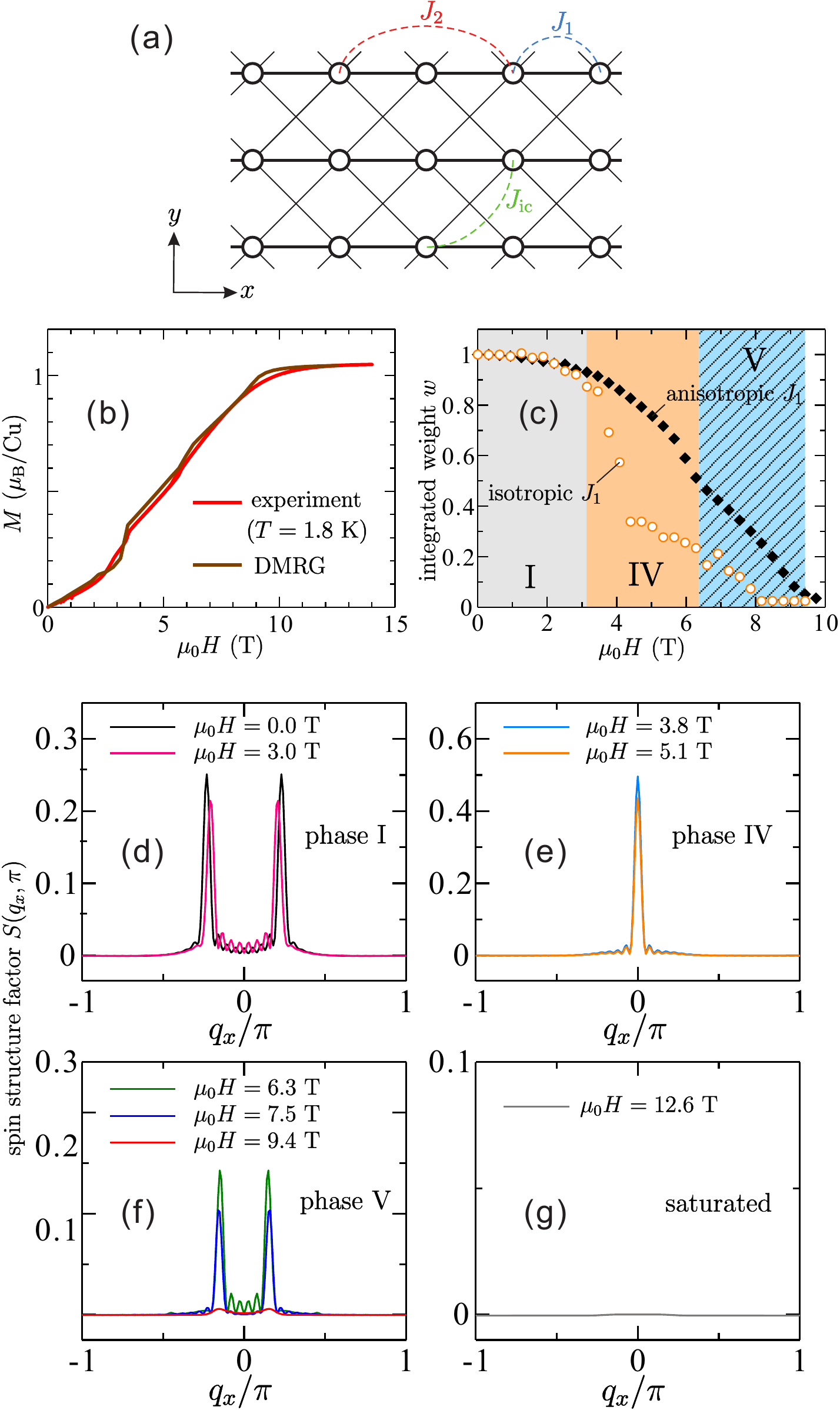}
\caption{(a) Cluster used in our DMRG calculations. (b) Calculated magnetization curve by DMRG method with $(J_1^x, J_1^y, J_1^z) = (-91.1, -86.6, -88.4)$, $J_2 = 28.3$ ($J_2/|\tilde{J}_1| \sim 0.32$), and $J_{\rm ic} = 2.7$. (c) Integrated weight of $S(q_x, \pi)$ as a function of field ($H \parallel b$), where the results using anisotropic and isotropic NN exchange couplings are compared. For the isotropic case we set $J_1^x = J_1^y = J_1^z = -88.4$ K. (d)--(g) Field dependence of static spin structure factor $S(q_x,\pi)$.}
\label{fig_DMRG}
\end{center}
\end{figure}

To provide a first reasonable explanation for the experimentally observed magnetic phases and magnetization, we introduce frustrated $XYZ$ Heisenberg chains coupled by diagonal interchain exchange interaction (see Fig.~\ref{fig_DMRG} (a)). The Hamiltonian is given by
\begin{eqnarray}
\nonumber \mathcal{H} &=& \sum_{ij,\gamma=x,y,z} J_1^\gamma S^\gamma_{i,j}S^\gamma_{i+1,j} + J_2 \sum_{ij} \mathbf{S}_{i,j} \cdot \mathbf{S}_{i+2,j} \\
&+& J_{\rm ic} \sum_{ij} \mathbf{S}_{i,j} \cdot \mathbf{S}_{i+1,j\pm1} + h\sum_{ij} S_{i,j}^z,
\label{hamxyz}
\end{eqnarray}
where $J_1^\gamma$ and $J_2$ are the NN FM and the NNN AFM intrachain exchange couplings, $J_{\rm ic}$ is the diagonal interchain exchange coupling, and $S^\gamma_{i,j}$ is the $\gamma$-component of spin-operator $\mathbf{S}_{i,j}$ at $i$-th site on $j$-th chain.

Fig.~\ref{fig_DMRG} (b) shows the magnetization $M$ curve measured at $T = 1.8$\,K, where the external magnetic field $H$ is applied along the $b$ axis. In spite of a fairly low experimental temperature that should prevent a significant finite-temperature effect, the magnetization saturates only asymptotically in the limit $T = 0$\,K with increasing $H$ following a power-law $\propto 1/H^2$ in the limit $H \rightarrow \infty$~\cite{Schumann2018}. Such an unusual behavior reflects the presence of strong quantum fluctuations. It is in sharp contrast with the behavior of typical isotropic 1D and 2D spin systems which exhibit a divergent increase of $M$ near the saturation at very low temperature. This asymptotical saturation of $M$ can be explained by assuming an $XYZ$ anisotropy of the NN intrachain exchange coupling $J_1$~\cite{grafe2017,Schumann2018}. 

To estimate a possible parameter set, we performed an approximate fit of the experimental data using the density-matrix renormalization group method with cluster size $L_x \times L_y = 40 \times 4$. Note that the DMRG results shown in this paper are for zero temperature. As shown in Fig.~\ref{fig_DMRG} (b), we have found an optimal description of the experimental data by setting $(J_1^x, J_1^y, J_1^z)=(-91.1, -86.6, -88.4)$, $J_2 = 28.3$ ($J_2/|\tilde{J}_1| \sim 0.32$), and $J_{\rm ic} = 2.7$ in units of K. These values reasonably agree with those estimated by fitting the inelastic neutron scattering data~\cite{rule2017}.

In this context, we note that Eqs.~(\ref{eq_T1}) and (\ref{eq_T1beta}) might be refined/generalized by adding the asymptotical field dependence skipped for the sake of simplicity. The presence of a staggered DM vector {\it \bf D} $\perp$ to the $b$ axis allowed by the monoclinic symmetry of linarite may affect the field induced local transverse polarization which also exhibits an asymptotical power-law and this way also the spin-gap and the stretching behavior considered in Eq.~(\ref{eq_T1beta}). A more detailed quantitative consideration of this difficult problem is postponed to a future investigation of the corresponding $XYZ$+DM-model.

In order to investigate the magnetic structure, we calculated the static spin structure factor defined by
\begin{equation}
S(q_x, q_y) = \frac{1}{L_x L_y}\sum_{i j k l}\langle \mathbf{S}_{i,j} \mathbf{S}_{k,l} \rangle \exp[i\mathbf{q}\cdot(\mathbf{r}_{i,j} - \mathbf{r}_{k,l})],
\end{equation}
where $\mathbf{r}_{i,j}$ is the position of the spin at the $i$-th site on the $j$-th chain. We could make a rough assessment of the observed magnetic moment by integrating the structure factor $S(q_x,\pi)$ over $q_x$, i.e., the integrated weight $w=\int_{-\pi}^\pi S(q_x,\pi)dq_x$. In Fig.~\ref{fig_DMRG} (c) $w$ is plotted as a function of $\mu_0 H$. Assuming  $XYZ$ anisotropy of NN coupling, only, the tendency seems to be qualitatively in accord with the value of the magnetic moment measured by neutron diffraction at $T$\,$=$\,0.1\,K. This result provides further support for the importance of $XYZ$ exchange anisotropy in linarite.

From the spin structure factor $S(q_x,\pi)$ shown in Fig.~\ref{fig_DMRG} (d)--(g), we found four different magnetic phases as a function of the field $\mu_0 H$, and which are also visible as fine structure in the calculated magnetization: At 0\,T\,$\le$\,$\mu_0 H$\,$\lesssim$\,3.2\,T two peaks are seen at $q_x = \pm q_{x,{\rm incomm}}$ indicating an incommensurate ordering along the chain direction (phase I). The intensity of the peaks and the value of $q_{x,{\rm incomm}}$ are slightly reduced by $\mu_0 H$. At 3.2\,T\,$\lesssim$\,$\mu_0 H$\,$\lesssim$\,6.1\,T a single peak appears at $q_x = 0$, which corresponds to a commensurate AFM spin alignment (phase IV). This is consistent with the neutron diffraction scans at low temperature (see Fig.~\ref{fig:Overview_kscans_D10_7p5T}). At 6.1\,T\,$\lesssim$\,$\mu_0 H$\,$\lesssim$\,9.6\,T, surprisingly, incommensurate peaks are re-emergent, in line with the spin-density wave phase in phase V. The peak position roughly agrees with the predicted periodicity of the spin-density wave with inherent nematic correlations: $q_x / \pi = (1 - M/M_{\rm s})/2$, where $M_{\rm s}$ is the saturation magnetization. In fact, a nematic (two-magnon bound type) state at high magnetization has been suggested in the presence of $XYZ$ exchange anisotropy~\cite{grafe2017}. At $\mu_0 H$\,$\gtrsim$\,9.6\,T the spins are (almost) fully saturated. The calculated phase boundaries somewhat deviate from the observed ones. Perhaps, the DM couplings and further smaller exchange interactions should be taken into account for more quantitative considerations.

\section{Discussion}

From our present study, and taking account of recent investigations on linarite~\cite{Cemal2018,Feng2018}, the experimental case of the phase diagram appears to be even more complex than believed so far. In the Refs.~\cite{Cemal2018,Feng2018} it has been demonstrated that in order to understand linarite it will be necessary to tackle the issue of magnetic anisotropy. At present, it still has not been finally resolved what role such aspects as exchange anisotropy, (staggering of) the $g$ tensor or the Dzyaloshinsky-Moriya interaction play with respect to linarite in particular, and for the frustrated $J_1$-$J_2$ spin chain in general. But even if we limit our view to a particular crystallographic direction of linarite -- like in our study the $b$ axis -- again and again new features and anomalies show up upon lowering temperature and increasing the magnetic field.

\begin{figure}[t!]
\begin{center}  
\includegraphics[width=0.95\columnwidth]{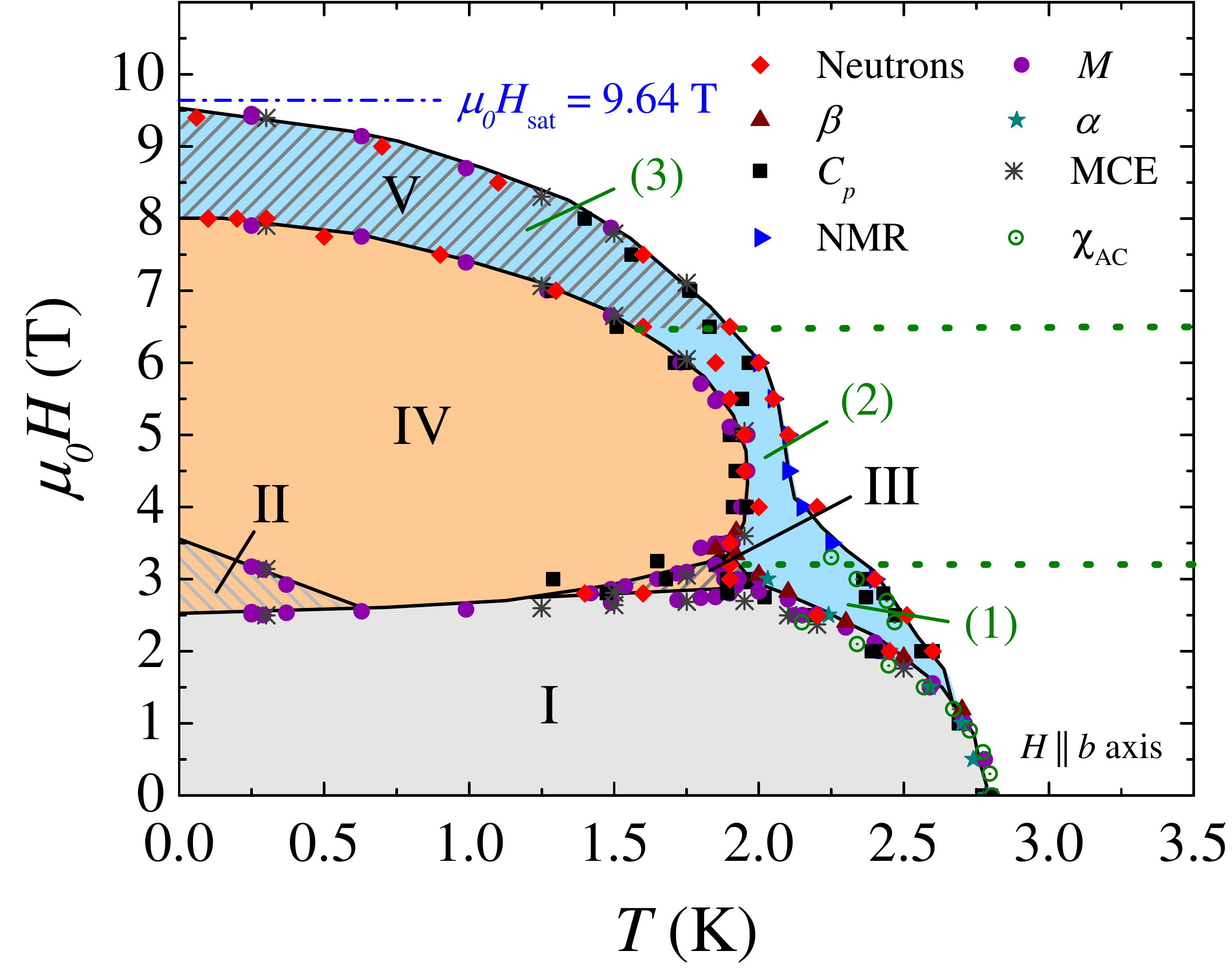}
\end{center}
\caption{Updated magnetic phase diagram of linarite for $H \parallel b$ axis. The red diamonds represent the ($T$, $\mu_0 H$) points where a change of the magnetic behavior is observed in the neutron diffraction experiments. The solid black lines representing the phase borderlines were drawn from the thermodynamic measurements. The green dotted lines highlight the magnetic fields where a change of the behavior of the incommensurability vector $\vec{q}$ = (0 $k_y$ 0.5) in phase V has been observed in neutron diffraction measurements (see Fig.~\ref{fig:PropagationVector}). These regions are therefore named phase V (1), (2) and (3).} 
\label{fig:PhaseDiagram}
\end{figure}

As we already noted, the case made in Ref.~\cite{Cemal2018} for a particular field dependence of the incommensurability $k_y$ in phase V for fields $H \parallel b$ axis is incomplete. Our data do not support the scenario of a slight increase of $k_y$ with field at lowest temperatures, while the anomalous field/temperature dependence in low and intermediate fields is unaccounted for. On a qualitative level, it appears as if phase V is governed by various and coexisting but competing sub-components. In result, it leads to a situation where phase V is  probably not a magnetically homogeneous phase in the same way as the other phases I and IV but may contain more or less hidden "sub-phases" or different regimes separated by crossovers. Correspondingly, in Fig.~\ref{fig:PhaseDiagram} we draw a modified phase diagram for $H \parallel b$ axis. In the figure we include the transition temperatures/fields obtained from neutron diffraction. For the borderlines from phases I and III the neutron data points are in good agreement with those obtained by other techniques. For the borderline between IV and V there is some scatter, which likely reflects the hysteretic first order nature of this transition, with a narrow coexistence region of phases IV and V.

In addition, in phase V we distinguish three different regions: (1) The low field region ($\mu_0 H$\,$<$\,3.2\,T) is defined by a decreasing $k_y$ for increasing temperature. As well, the transition V--PM shifts down to $\sim 2.4$\,K. (2) At intermediate fields (3.2 to 6.5\,T) $k_y$ increases with increasing temperature, while the transition V--PM shifts even further down to $\sim 2.2$\,K and shows hardly any field dependence between 4 and 6\,T. From 6\,T on the upper transition temperature is further pushed towards lower temperatures. (3) In the high field region of the magnetic phase diagram ($\mu_0 H > 6.5$\,T), phase V is phase separated into an incommensurate magnetic component with $k_y (\mu_0 H, T)$ without a significant temperature or field dependence and a component of unknown microscopic nature. The transition V--PM is suppressed to zero with magnetic field.

\begin{figure}[b!]
\begin{center}
\includegraphics[width=0.98\linewidth]{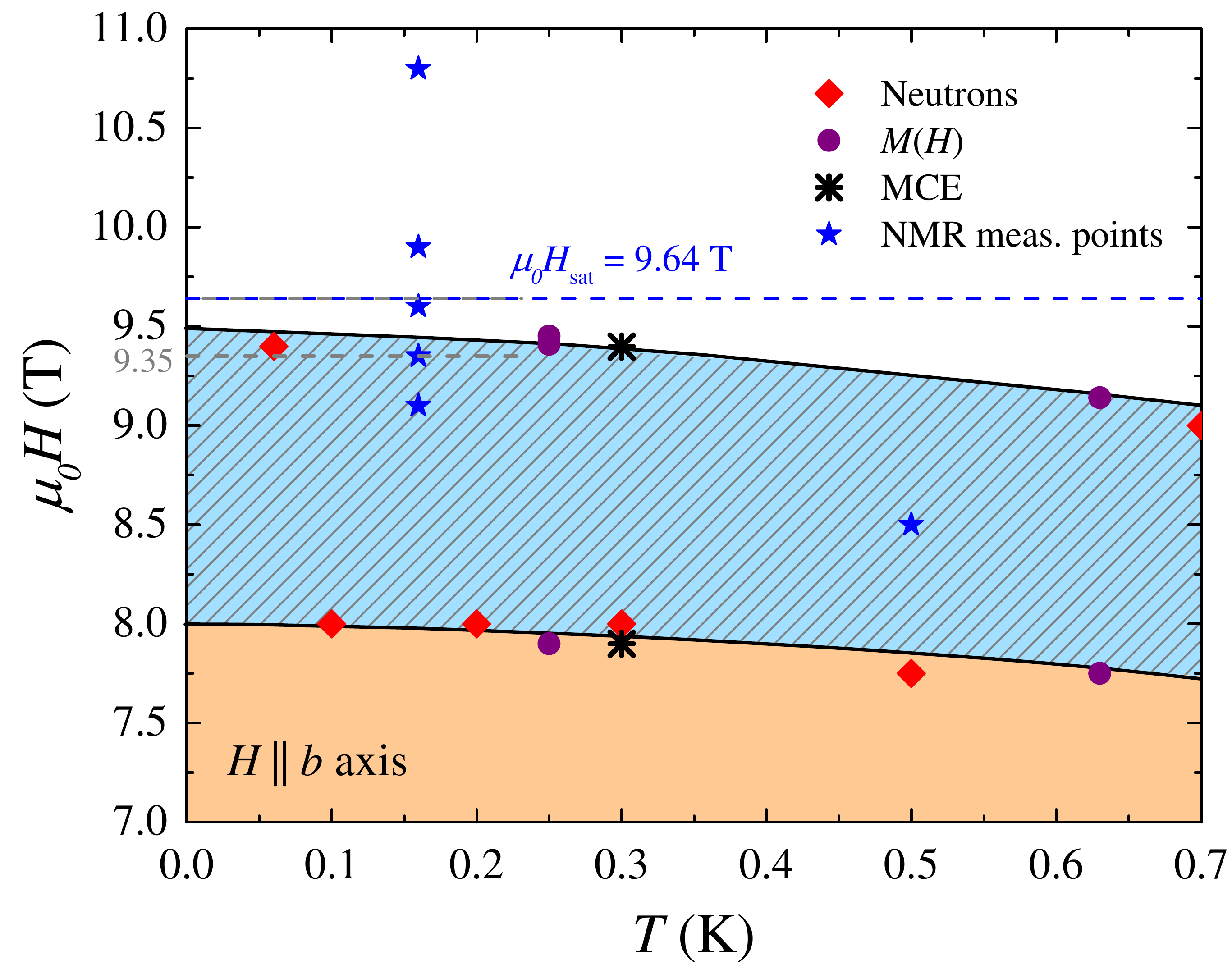}
\caption{Enlarged high-field section of the updated magnetic phase diagram of linarite for $H \parallel b$ axis; labels as in Fig.~\ref{fig:PhaseDiagram}. The blue stars indicate the points within the magnetic phase diagram where the NMR measurements from Fig.~\ref{NMRspectrum} have been carried out. The solid black lines representing the phase borderlines were drawn from the thermodynamic measurements. The gray line indicates the lower border for a possible nematic state as observed by NMR.} 
\label{PhaseDiagram_hf}
\end{center}
\end{figure}

Especially, the low temperature/high field section from phase V to magnetic quasi-saturation exhibits an even more intricate behavior. To illustrate this, in Fig.~\ref{PhaseDiagram_hf} we zoom into the corresponding phase region. In the figure, we include the thermodynamic data points signaling a phase transition, the neutron diffraction data points indicating the borderline between phases IV and V, as well as the suppression of phase V magnetic order, and the positions of the present NMR experiments in the phase diagram. Moreover, we include the position of the quasi-saturation field $\mu_0 H_{\rm sat} = 9.64$\,T.

From the figure, we find a good matching of the phase borderline IV--V from thermodynamic techniques and neutron diffraction. Evidently, the disappearance of commensurate magnetic order is equally well reflected in the neutron diffraction experiment as by thermodynamic probes. For the upper borderline of phase V, at least below $\sim$\,300\,mK the neutron diffraction data are slightly lower than the magnetocaloric/magnetization data. From the data, it is unclear if this just reflects an experimental uncertainty or if it is an intrinsic feature. Notably, however, the upper boundary of phase V for $T \to 0$\,K obtained either from thermodynamics (9.5\,T) or neutron diffraction (9.4\,T) is distinctly lower than the quasi-saturation field determined from NMR. Therefore, there is a finite high field/low temperature phase range without long-range dipolar magnetic order.

In terms of our NMR study, in fact the regime 9.1\,T $\leq \mu_0 H \leq$ 9.64\,T is a rather peculiar one: The broad distribution of the NMR signal at 9.1\,T and also at 9.35\,T would imply a broad local field distribution static on the time scale of NMR. This is neither in accordance with a strongly field-polarized paramagnetic phase nor with a (quasi-)saturated or SDW state. The evolution of $1/T_1$ shows that the lowest energy excitation in the saturated regime $\mu_0 H \geq 9.64$\,T are two-magnon excitations pointing possibly to the formation of bound-magnon pairs. The condensation of these bound-magnon pairs into a spin quadrupolar order may occur within the field interval 9.35\,T\,$\leq$\,$\mu_0 H$\,$\leq$\,9.64\,T, where a relatively narrow NMR line was observed. Only, the temperature dependence of the relaxation rate in this field interval shows a rather unusual character and neither proves nor disproves the realization of a spin multipolar state. 

The presence of a temperature dependent NMR signal in the low temperature/high field regime might indicate the presence of a coexisting competing dipolar component,  pointing to a crossover region or to a specific dipolar phase directly related to hidden multipolar order in the spirit of Ref.~\cite{smerald2016}. Thermal excitations over a highly anisotropic or even nodal one- or two magnon gap might also cause unusual $T$-dependencies. From this we conclude that a narrow "nematic" phase, called phase VI (or a region with dominant, strongly field induced nematic correlations) is nearby. The identification of a spin multipolar state from the field and temperature dependence of the NMR relaxation rate is not straightforward because the physical properties of linarite change rapidly with field just below the magnetic quasi-saturation and near or slightly above the inflection point of the longitudinal magnetization. Measurements (like Raman scattering) which directly probe the tensorial character of the supposed corresponding nematic order parameter would be helpful to elucidate this challenging issue. Similarly, a novel resonance mode to be detected in ESR or INS studies in this experimentally difficult-to-access region at very low temperature and relatively high fields, would be helpful to resolve this puzzle. In addition, theoretical studies on a more realistic model, taking for example into account Dzyaloshinsky-Moriya interactions or interchain coupling are also needed to confirm the prediction of a single magnon spin gap in such anisotropic $J_1$-$J_2$ Heisenberg chain systems.

In summary, for linarite with $H \parallel b$ axis, the experimental case of the magnetic phase diagram appears to be far more complex than initially proposed. From neutron diffraction and NMR measurements, there appears to be a field range just below quasi-saturation ($\sim$ 9.1 -- 9.64\,T) containing rather unusual physics. If this can be associated to multipolar states will have to be further tested in the future.

\begin{acknowledgments}
We would like to thank the ILL, the HZB and ANSTO for the allocation of neutron radiation beamtime. We acknowledge fruitful discussion with Hans-Joachim Grafe, Cli\`{o} Efthimia Agrapidis, Johannes Richter, Rolf Schumann, Andrey Zvyagin, and Alexander Tsirlin. This project has received funding from the European Union's Seventh Framework Programme for research, technological development and demonstration under the NMI3-II Grant number 283883. Our work has been supported by the DFG under Contracts WO 1532/3-2 and SU 229/9-2 and the SFB 1143.
\end{acknowledgments}


\begin{thebibliography}{00}
\bibitem{remsat} In low-symmetry systems with spin anisotropy of the $XYZ$-type and/or the presence of DM-couplings there is only asymptotical saturation for any field direction. The quasi-saturation field can be defined as the field where the highest inflection point in the longitudinal magnetization curves extrapolated to $T = 0$ or the maximum of transverse magnetization are observed. Saturation can be reached asymptotically, only. However, for weak exchange anisotropy an almost full magnetization is achieved with very small relative deviations of the order of 10$^{-4}$ for the field denoted as quasi-saturation in the text.
\bibitem{note1} Note that the gap value for this field, {\it i.e.}, $\Delta \approx 0.7$\,K, is not bigger than the temperature of the fit, yielding a less precise value for $\Delta$ for this field than for higher fields.
\end{thebibliography}
\end{document}